\documentclass[a4paper,onecolumn,11pt, accepted=2024-04-11, titlepage]{quantumarticle}
\pdfoutput=1
\usepackage[utf8]{inputenc}
\usepackage[english]{babel}
\usepackage[T1]{fontenc}
\usepackage{amsmath}
\usepackage{amssymb}
\usepackage{amsthm}
\usepackage{hyperref}
\usepackage{physics}
\usepackage{float}
\usepackage{graphicx}
\usepackage{ulem}
\usepackage{multirow}
\usepackage{xfrac}
\graphicspath{ {figures/} }
\usepackage{ulem}

\usepackage[numbers, sort&compress]{natbib}

\usepackage{tikz}
\usepackage{lipsum}

\usepackage{color}

\newcommand{\rd}[1]{{\color{black}#1}}

\newtheorem{theorem}{Theorem}

\begin{document}

\title{
  Minimal-noise estimation of noncommuting rotations of a spin}
	\author[0,1]{Jakub Czartowski}
	\email{jakub.czartowski@doctoral.uj.edu.pl}
	\orcid{0000-0003-4062-833X}
	\author[1,2,3]{Karol Życzkowski}
	\orcid{0000-0002-0653-3639}
	\author[4]{Daniel Braun}
	\affil[0]{Doctoral School of Exact and Natural Sciences, Jagiellonian University, ul. Łojasiewicza 11, 30-348 Kraków, Poland}
	\affil[1]{Faculty of Physics, Astronomy and Applied Computer Science, Jagiellonian University, ul. Łojasiewicza 11, 30-348 Kraków, Poland}
	\affil[2]{Centrum Fizyki Teoretycznej PAN, Al. Lotników 32/46, 02-668 Warszawa, Poland}
	\affil[3]{National Quantum Information Center (KCIK), University of Gda{\'n}sk,
		Poland}
	\affil[4]{Institute of Theoretical Physics,	University of T{\"u}bingen,
		Auf der Morgenstelle 14, 72076 T{\"u}bingen, Germany}
	\bigskip
	
	\date{November 30, 2023}
	
	\maketitle
	
	\begin{abstract} 
		We propose an analogue of  $\text{SU}(1,1)$ interferometry to measure 
		rotation of a spin by using two-spin squeezed states.
                Attainability of the Heisenberg limit %
                for the estimation of the rotation angle %
                is demonstrated for maximal squeezing.
                For a   specific
                squeezing     direction and strength 
     an 
     advantage in sensitivity for {\it all} %
     equatorial rotation axes  (and hence non-commuting rotations)
     over the classical bound is shown in terms of quadratic scaling of 
     the single-parameter quantum Fisher information for the corresponding rotation angles. Our results provide a method for measuring magnetic fields in any direction in the $x$-$y$-plane with the same optimized initial state.  

	\end{abstract}
%
%
	
	\section{Introduction} \label{sec:intro}
	

		For more than a century light interferometry has proven 
		to be an invaluable tool for the progress of physics. 
		Recent developments of $SU(1,1)$ interferometry making use of squeezed states of light are applied at the cutting edge of experimental physics of gravitational waves and axions  \cite{DBS13, AEtAl13, IR18, BEtAl19}.  

			
        In a seminal 
        work Yurke, McCall and Klauder \cite{YMK86} considered a Mach-Zehnder interferometer with the beam-splitters replaced by non-linear crystals as four-wave mixers, pumped with a common strong laser beam. %
        While in the standard Mach-Zehnder interferometer the variance of 
        any unbiased estimation of the phase shift $\phi$ in one arm relative to the other scales with the number $N$ of photons as the inverse square root, $\sqrt{\Delta\phi^2 }\propto O(N^{-1/2})$, 
        the so-called standard quantum 
        limit, they demonstrated that 
        the modified interferometer  
        generates a $\text{SU}(1,1)$ transformation of the annihilation and creation operators of the two modes and %
        achieves the
         scaling $\sqrt{\Delta\phi^2 }\propto O(N^{-1})$, 
        now known as the Heisenberg limit. In a recent reinterpretation, Caves \cite{Caves20}         argued 
        that such a $\text{SU}(1,1)$ interferometer can be understood as a ``displacement detector'': On an initial coherent state it generates a sequence of squeeze-displace-unsqueeze operations, where the displacement refers to the displacement of the label of the coherent state in the complex plane.  The net result of the three operations is a noiseless amplification of the displacement. 
         
     Using two oscillators, the added noise in the displacement of {\sl both quadratures} can be made arbitrarily small \cite{PhysRevX.2.031016}. The noiseless amplification of displacements of two non-commuting quadratures should be contrasted with the fact that two non-commuting quadratures themselves cannot be measured simultaneously with arbitrarily small statistical error, as this would violate Heisenberg's uncertainty relation. Nor is it possible to amplify even a single quadrature without adding noise (see e.g.~\cite{AgarwalQOpticsBook}, p.220 ff), such that the quantumness of an amplified signal is normally quickly lost in the amplification process.  Measuring the  displacement of both quadratures
    is achieved by using two modes, one for each quadrature, whose relative coordinate and total momentum (in oscillator parlance) {\sl do} commute, hence allowing arbitrarily small statistical error for both displacements.


        In the microwave domain, noiseless amplification of small signals plays an important  role and has been boosted by the tremendous technological progreess in the context of quantum computing based on super-conducting qubits \cite{bergeal_phase-preserving_2010,PhysRevB.98.045405,PhysRevLett.106.110502,renger_beyond_2021}, as well as with nano-mechanical oscillators \cite{PhysRevLett.118.103601}.  Recently, noiseless microwave amplification has even been proposed for speeding-up the search for axions \cite{Sik83,zheng_accelerating_2016}.   
        


Here we transfer the SU(1,1) protocol to spin-systems.  The displacement of a coherent state of a harmonic oscillator has its analogue in the rotation of spin-coherent states, which, in the case of physical spins, can be generated with applied magnetic fields (whereas pseudo-spins-$j$ representing generic $2j$-level systems can be rotated by driving transitions between these levels). Our investigations have therefore direct application for magnetometry, where one wants to measure magnetic fields in different directions.          
We follow the interpretation of Caves of SU(1,1) interferometry and 
transplant it to the realm of spins by employing the spin squeezed states introduced by Kitagawa and Ueda in \cite{KU93} and their extension to 2-spins squeezed states \cite{Byr13, KPST13}. 
%
Schemes utilising squeezing of spin states have been recently demonstrated experimentally to yield significant advantages over the classical limits of sensitivity in the context of Bose-Einstein condensates \cite{SDC11, CGSP21}, ultracold dipolar molecules \cite{BEtAl21}, spectroscopy on beryllium atoms \cite{LEtAl04}, rotosensors \cite{CEtAl21}, programmable quantum sensors \cite{KEtAl19}, and others \cite{REtAl10, CEtAl22, KSVZ23}.

We demonstrate that a squeeze-rotate-unsqueeze (SRU) protocol for a
single spin as well as for a two-spin system is able to achieve the
Heisenberg scaling of 
the single-parameter quantum Fisher information 
(QFI) for rotation estimation, $\text{QFI}\propto N^2$ in the size of
the displaced system, for particular equatorial axes of
rotation. Moreover, for a particular value of squeezing we demonstrate
scaling of the single parameter 
$\text{QFI}\propto N^2/2$ for {\it any} equatorial rotation
axes, hence for non-commuting rotations. Both results are achieved
using oversqueezing, thus demonstrating its usefulness for the quantum
metrological applications. Finally, we consider the problem of
estimation of the rotation axis, as contrasted with the estimation of
the rotation angle around a known axis. \rd{Therein we show that 
  the quantum Cramer-Rao bounds are attainable given a proper choice
  of the 
  parameter values 
  for which the estimation is carried out. 
} 
		

A word is in order about the possibility to estimate the rotation angles
for non-commuting rotations (with rotation axes in the $xy$-plane)
``simultaneously'' with Heisenberg-limited sensitivity. In the
framework of multi-parameter quantum metrology, with ``simultaneously'' it is
commonly understood that there exists a measurement that saturates the
matrix-valued quantum Cram\'er-Rao 
bound for the co-variance matrix of the estimators.  It requires the
commutation on average of the corresponding symmetric logarithmic
derivatives (SLDs) of the density matrix. As we shall see,
this is, for the problem at hand, only possible for few, specific
combinations of parameters. 
Rather, what we show generically, is that the values of the single-parameter QFI for
different rotation angles about rotation axes in the $xy$-plane that
lead to non-commuting rotations all scale $\propto N^2$ up to
exponentially small terms in probe spin $S_P$.  This is  
exactly in the same spirit as noiseless amplification of 
both quadratures of a single mode of an electro-magnetic field. 
To illustrate this, we provide in Appendix \ref{app:SDU_QFI} a general
analysis of the   SU(1,1) interferometry and the SDU protocol using the lens of
  QFI. In case of a single-mode SDU protocol it proves 
  that only one direction of the displacement at a time can be
  estimated with an enhanced sensitivity. For a two-mode protocol
  serving as a basis for the SU(1,1) interferometry we show that the QFI
  matrix is proportional to the identity.  This allows estimation
  of displacements in any direction with the same enhanced sensitivity.
  Nevertheless, the symmetric logarithmic derivatives do not commute, which prevents any
  multivariate estimation scheme to 
  saturate the multi-parameter quantum Cramer-Rao bound for
  simultaneous estimation of the displacement of the two quadratures.

	\section{Setting the Scene} \label{sec:scene}
	

	\subsection{Group-theoretic approach to interferometry}
	
		Consider a standard Mach-Zehnder interferometer consisting of two arms and describe the state of light in its two arms by annihilation operators $\hat{a}$ and $\hat{b}$ respectively, together with their standard commutation relations $\comm{\hat{a}}{\hat{b}} = \comm{\hat{a}^\dagger}{\hat{b}^\dagger} = 0$ and $\comm{\hat{a}}{\hat{a}^\dagger} = \comm{\hat{b}}{\hat{b}^\dagger} = \mathbb{I}$.	It is then easily verified that operators of the form
		\begin{align}
			\hat{J}_1 = \frac{1}{2}\qty(\hat{a}^\dagger \hat{b} + \hat{b}^\dagger \hat{a}), &&
			\hat{J}_2= -\frac{i}{2}\qty(\hat{a}^\dagger \hat{b} - \hat{b}^\dagger \hat{a}), &&
			\hat{J}_3 = \frac{1}{2}\qty(\hat{a}^\dagger \hat{a} - \hat{b}^\dagger \hat{b}),
		\end{align}
		together with the number operator $\hat{N} =\hat{a}^\dagger \hat{a} + \hat{b}^\dagger \hat{b}$  provide an SU(2) representation with the Casimir operator given by $\hat{J}^2 = \frac{\hat{N}}{2}\qty(\frac{\hat{N}}{2} + 1)$. Making use of the isomorphism $\text{SU}(2)\cong\text{SO}(3)$ and the resulting geometric interpretation, Yurke \textit{et al.} have shown that the standard limit of fluctuation of phase estimation is given as $\Delta\phi = N^{-1/2}$ with $N = \hat{\ev{N}}$ the average number of photons in a coherent state used in the interferometer \cite{YMK86}.
		
		In a similar manner they demonstrated that a four-wave mixer can be identified with a representation of a non-compact group SU(1,1) given by
		\begin{align}
			\hat{K}_0 = \frac{1}{2}\qty(\hat{a}^\dagger \hat{a} + \hat{b} \hat{b}^\dagger), &&
			\hat{K}_1 = \frac{1}{2}\qty(\hat{a}^\dagger \hat{b}^\dagger + \hat{b} \hat{a}), &&
			\hat{K}_2= -\frac{i}{2}\qty(\hat{a}^\dagger \hat{b}^\dagger - \hat{b} \hat{a}),
		\end{align}
		which provide generators for a set of squeezing operators. It has been demonstrated that this kind of active optical elements, when incorporated into the structure of Mach-Zehnder interferometer, allows for measurement of a phase shift with fluctuations on the order of $\Delta\phi\propto N^{-1}$.

		Caves in his reinterpretation \cite{Caves20} presents a useful perspective on the action of SU(1,1) interferometer as a composition of three operations -- squeezing, displacement, and unsqueezing (SDU in short), defined for a single mode as
		\begin{equation}
			\ket{\text{SDU}_1(\gamma; r)} = \hat{S}_1(r) \hat{D}(\gamma) \hat{S}_1^\dagger(r) \ket{0},
		\end{equation}
		where the squeezing operators are defined as $\hat{S}_1(r) = \exp(r(\hat{a}^2 - \hat{a}^{\dagger2})/2)$. The displacement operator reads,  $\hat{D}(\gamma) = \exp(\gamma \hat{b}^\dagger - \gamma^* \hat{b})$, and $\ket{0}$ is a vacuum state defined as a coherent state of light with the average of both quadratures equal to $0$. Similarly, a two-mode protocol is defined by a composition
		\begin{equation}
			\ket{\text{SDU}_2(\gamma; r)} = \hat{S}_2(r) \hat{D}(\gamma) \hat{S}_2^\dagger(r) \ket{0}\otimes\ket{0},
		\end{equation} 
		with the two-mode squeezing operators $\hat{S}_2(r) = \exp(-2ir\hat{K}_2)$, and single mode displacement $\hat{D}(\gamma)$.

		\begin{figure}[H]
			\centering
			\includegraphics[width=.67\linewidth]{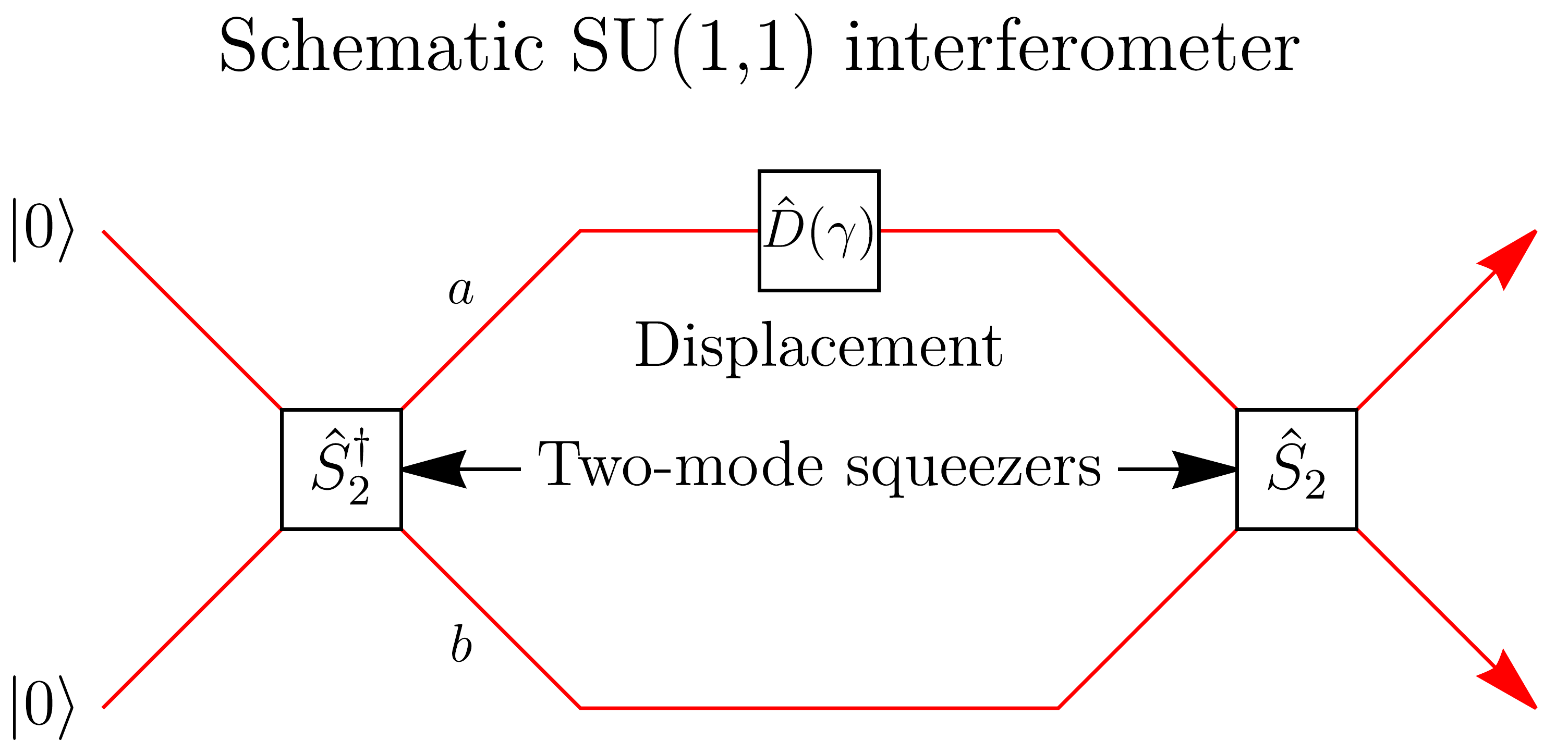} 
			\caption{
				Schematic depiction of an SU(1,1) interferometer with a displacement operator $\hat{D}(\gamma)$ only on one of its arms and the two-mode squeeze operators $\hat{S}_2 = \exp(-2ir\hat{K}_2)$.}
			\label{fig:SU11_interferometer}
		\end{figure}
		
		Caves demonstrated that for a single mode it is possible to achieve a resolution of $\ev{(\Delta\Re\gamma)^2} = e^{-2r}$ in an $n$-fold repetition of the protocol, at the cost of decreasing the complementary resolution, $\ev{(\Delta\Im\gamma)^2} = e^{2r}$. Furthermore, for the two-mode protocol he furthermore demonstrated the possibility of estimation of the full displacement with enhanced resolution, $\ev{(\Delta\gamma)^2} = e^{-2r}$. The possibility of achieving such an advantage for two non-commuting observables can be understood by noting that they can be reconstructed from measurements of displacements of two Bell variables $x_a + x_b$ and $p_a - p_b$, for which the squeezing occurs, while the other two variables are unsqueezed.
		
	\subsection{Spin Coherent and Squeezed States}
	
		Spin coherent states (SCS), first introduced in 1971
                by Radcliffe \cite{Rad71} by analogy to the coherent
                states of light and later analysed in further details
                by Arecchi \textit{et al.} and Holtz and Hanus
                \cite{ACGT72, HH74}, can be understood most easily in
                terms of a collection of $2S$ spin $\sfrac{1}{2}$
                particles, all in an eigenstate of the operator,
                $\sigma_{\theta,\phi} = \sin\theta\qty(\cos\phi\,
                \sigma_x + \sin\phi \,\sigma_y) + \cos\theta
                \sigma_z$, with the positive eigenvalue,
                $\sigma_{\theta,\phi} \ket{\theta,\phi}=
                \ket{\theta,\phi}$. By projecting the state
                $\ket{\theta,\phi}^{\otimes 2S}$ of such a collection
                onto the $2S+1$-dimensional subspace spanned by the
                symmetric states $\ket{S,M}$ one arrives at a well
                known expression, 
		\begin{equation}
			{\small\begin{aligned}
				\ket{S,S}_{\theta,\phi} & = P_S \ket{\theta,\phi}^{\otimes 2S} = \sum_{M=-S}^{S} \sqrt{\binom{2S}{S+M}}\cos[S+M](\frac{\theta}{2})e^{i(S-M)\phi}\sin[S-M](\frac{\theta}{2}) \ket{S,M},
			\end{aligned}}
		\end{equation}  
		where the basis states are defined as eigenstates of spin operator $S_z$, namely \mbox{$S_z \ket{S,M} = M \ket{S,M}$}, defining the basis for states constituting a $2S+1$-dimensional representation of the SU(2) group. The group-theoretic approach to the SCS and an extensive review can be found in \cite{ZFG90}.
	
\begin{figure}[H]
			\centering
			\includegraphics[width=.3\linewidth]{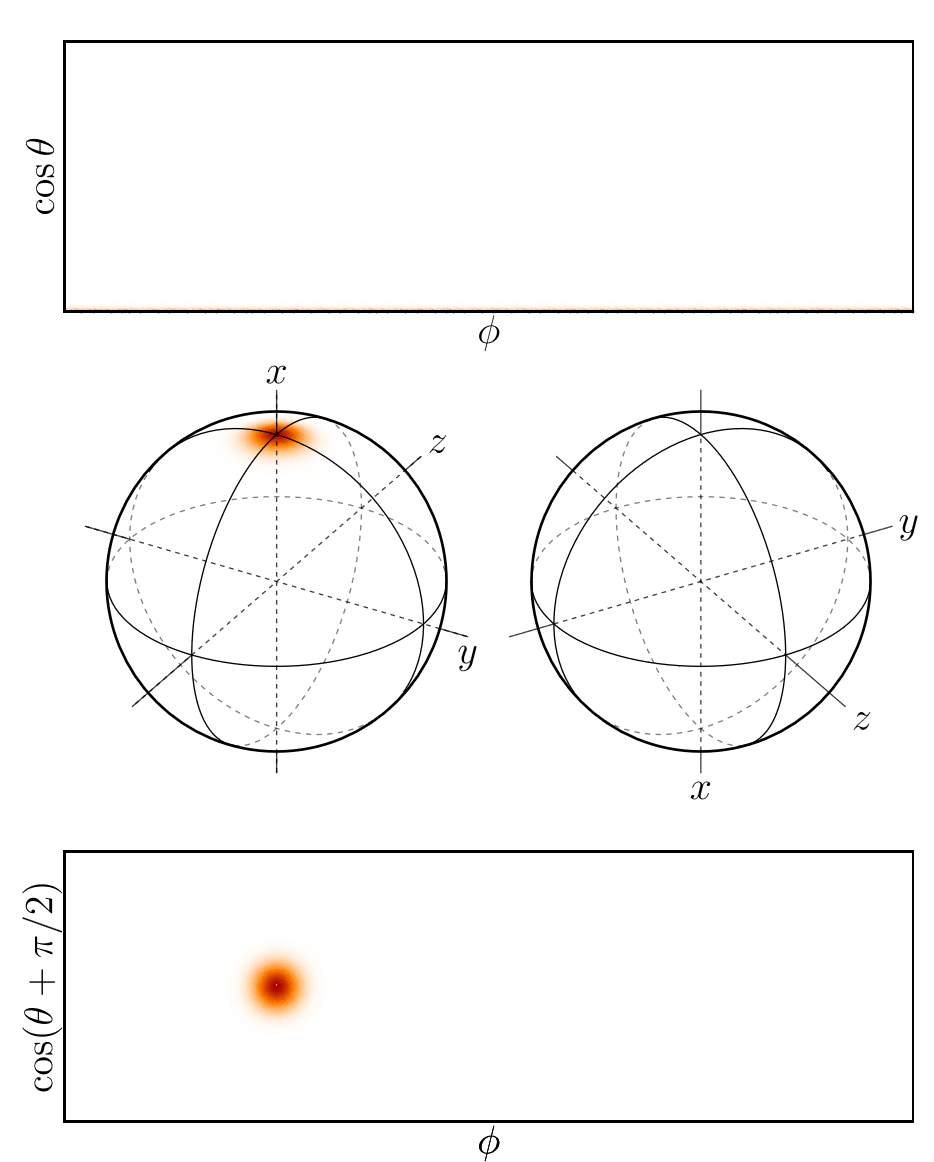} 
			\includegraphics[width=.3\linewidth]{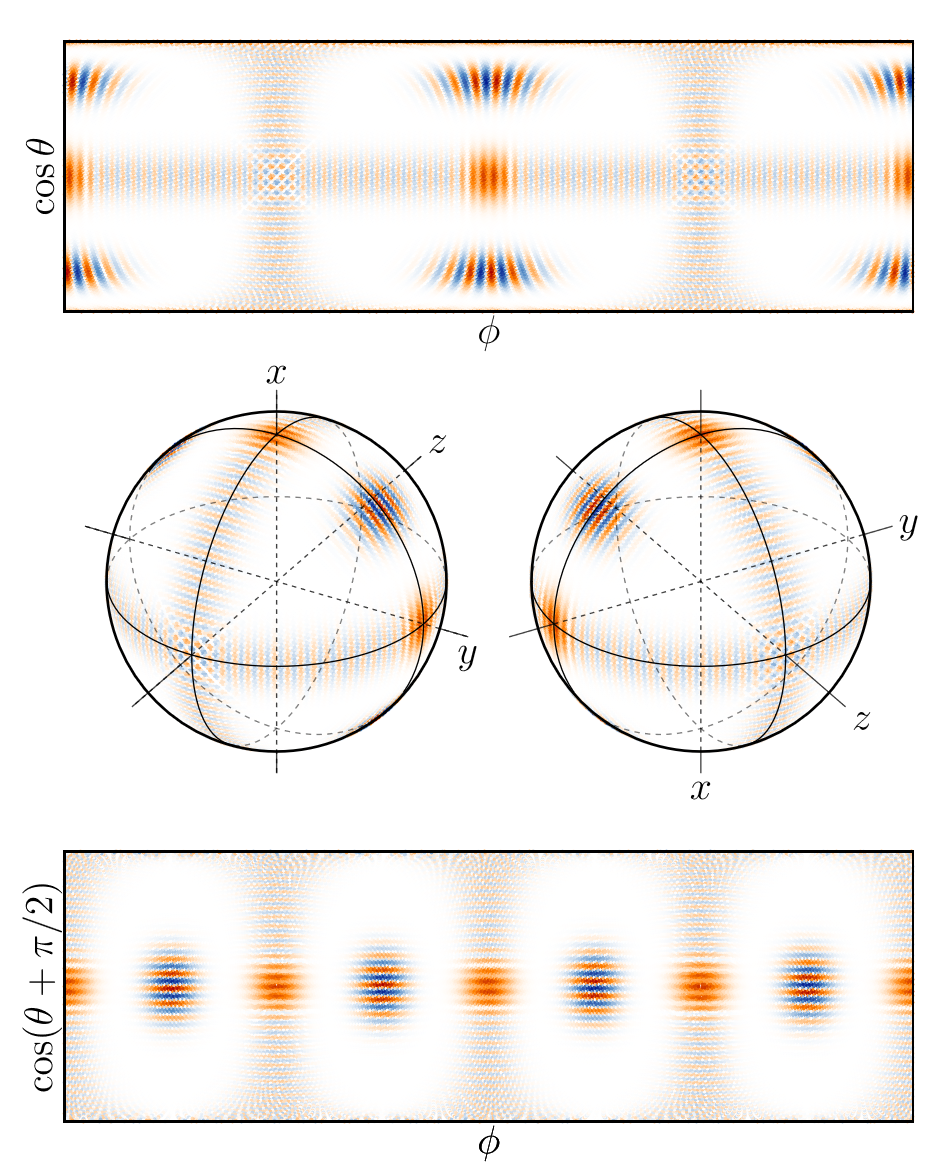} 
			\includegraphics[width=.3\linewidth]{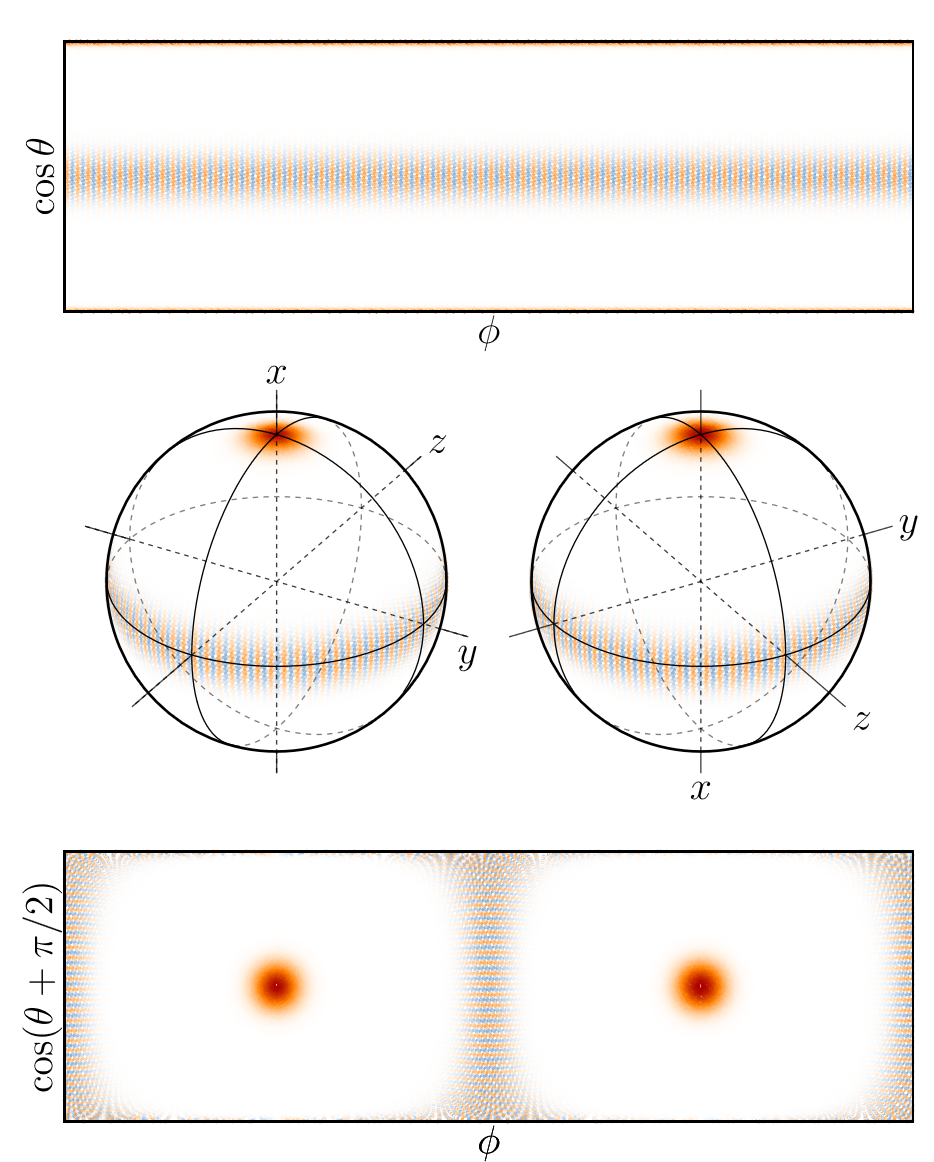} 
			\includegraphics[width=.06\linewidth]{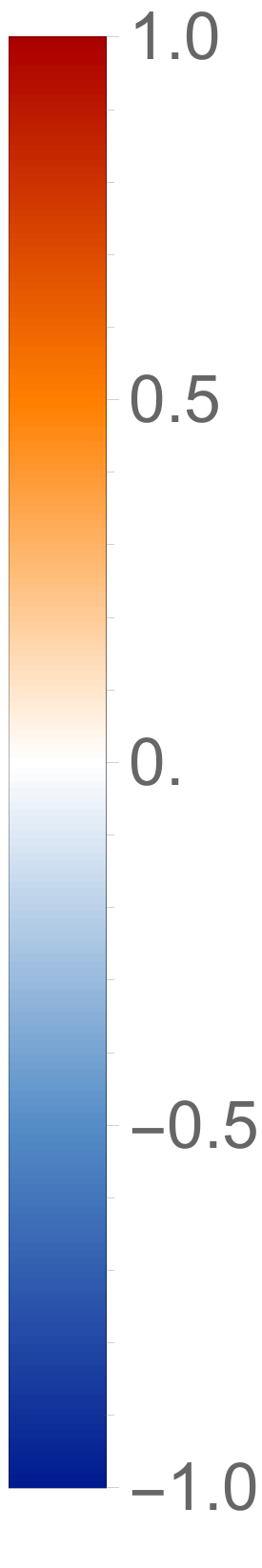}
			\caption{
			Spherical Wigner functions $W(\theta, \phi)$ for three examples of spin states with spin $S = 40$, of essential importance to the presented work. From left to right we present the spin coherent state (SCS), a squeezed spin state with intermediate squeezing $\mu = \pi/4$ and a maximally squeezed state with $\mu = \pi/2$. The states are given in four different projections; top and bottom projections are cylindrical with poles along $X$ and $Z$ axes, respectively. The middle ones are direct projections of the sphere with two different orientations indicated by the labels next to the axes.}
			\label{fig:wignershape}
		\end{figure}

		In a seminal work \cite{KU93} Kitagawa and Ueda reimagined what it means for a spin state to be squeezed by understanding that one should consider the squeezing in directions perpendicular to the main axis of a SCS. The authors proposed two squeezing hamiltonians,
		\begin{align}
			H_1 = \mu S_z^2, &&
			H_2 = -i\mu\qty(S_+^2 - S_-^2),
		\end{align}
		also referred to in the literature as one-axis squeezing (OAT) and two-axis counter-twisting (TACT), respectively. By evolving the SCS with each of them one gets one- and two-axis squeezed spin states, respectively. In this work we will be particularly interested in the former, which defines the squeezed states
		\begin{equation}
			\begin{aligned}
				\ket{S;\mu}_{\theta,\phi} & := e^{i \mu S_z^2}\ket{S,S}_{\theta,\phi} \\
				& = \sum_{M=-S}^{S} \sqrt{\binom{2S}{S+M}}\cos[S+M](\frac{\theta}{2})e^{i(S-M)\phi}\sin[S-M](\frac{\theta}{2})e^{i\mu M^2} \ket{S,M}.
			\end{aligned}
		\end{equation}
		In particular, we will be interested in squeezing the
                SCS along the $X$ axis, therefore we introduce the notation
		\begin{equation} \label{eq:squeezed_spin_sx}
			\ket{S;\mu}\equiv \ket{S;\mu}_{\sfrac{\pi}{2},0} = \frac{1}{2^S} \sum_{M=-S}^{S} \sqrt{\binom{2S}{S+M}} e^{i\mu M^2} \ket{S,M}.
		\end{equation}
		We note that
                $\ket{n,n;\pi}=\ket{n,n}_{\sfrac{\pi}{2},\pi}$ 
                and
                $\ket{\frac{2n+1}{2},\frac{2n+1}{2};\pi}=\ket{\frac{2n+1}{2},\frac{2n+1}{2}}_{\sfrac{\pi}{2},0}$.
                Note that squeezed spin states can be oversqueezed,
                unlike for the squeezed states of light, which has
                been emphasized in the original work of Kitagawa and
                Ueda. In Fig. \ref{fig:wignershape} we present three
                examples of squeezed spin states most important in
                this work, ranging from SCS with $\mu = 0$ to the
                maximum squeezing of $\mu = \sfrac{\pi}{2}$. For the
                visualization we go to the spherical Wigner function,
                as defined in \cite{Agar81}, later analysed in
                \cite{DAS94} and \cite{DKMG21} in the context of
                negativity of the spherical Wigner function $W(\theta,
                \phi)$ \cite{KZ04}. 
	
		A two-spin counterpart for one-spin squeezing has already been studied in experimental works \cite{Byr13, KPST13}, whereas the two-spin two-axis squeezing has been studied only recently in \cite{KEtAl20}, where its chaotic behaviour has been demonstrated. Therefore we restrict our attention in this work to the two-spin one-axis squeezed states defined as
		\begin{equation}
			\ket{S_M,S_P;\mu}_{\theta_M,\phi_M,\theta_P,\phi_P}:= e^{i \mu S_z \otimes S_z} \ket{S_M,S_M}_{\theta_M,\phi_M} \ket{S_P,S_P}_{\theta_P,\phi_P}.
		\end{equation}
	
	\subsection{Quantum Fisher Information and Hamiltonian Estimation}
		As a proxy for the sensitivity of rotation estimation
                we will use the quantum Fisher information in
                connection to the quantum Cram\'er-Rao bound, defined
                in the following theorem 
\cite{Hels67,braunstein_statistical_1994,braunstein_generalized_1996,Paris09,Fraise17}.
		\begin{theorem}
			Consider a family of states $\qty{\rho_\theta}$ depending on a scalar parameter $\theta\in\Theta$. The variance $\operatorname{Var}[\hat{\theta}_{\text{est}}]$ of any locally unbiased estimator $\hat{\theta}_{\text{est}}$
			\begin{align*}
				\eval{\operatorname{E}[\hat{\theta}_{\text{est}}]}_{\theta_0} = \theta_0, && 
				\dv{\theta}\eval{\operatorname{E}[\hat{\theta}_{\text{est}}]}_{\theta_0} = 1,
			\end{align*}
			is bounded from below by the inverse of the Quantum Fisher Information
			\begin{equation}
				\operatorname{Var}[\hat{\theta}] \geq \frac{1}{I_\theta(\rho_\theta)}
			\end{equation}
			with $I_\theta(\rho_\theta)$ the quantum Fisher information (QFI).
		\end{theorem}
	
		The QFI can be defined in terms of the symmetric logarithmic derivative (SLD) $L_{\rho_\theta}$,
		\begin{equation}
			I_\theta(\rho_\theta) := \tr[\hat{L}^2_{\rho_\theta} \rho_\theta],
		\end{equation}
		where the SLD itself is defined implicitly by
		\begin{equation*}
			\pdv{\rho_\theta}{\theta} = \frac{\hat{L}_{\rho_\theta} \rho_\theta + \rho_\theta \hat{L}_{\rho_\theta}}{2}.
		\end{equation*}
	
		For the purpose of this work, however, we 
                are mostly concerned with the closed form of QFI for pure states. Given that $\rho_\theta = \op{\psi_\theta}$ one finds that the expression for QFI reduces to 
		\begin{equation}
			I_\theta(\ket{\psi_\theta}) = 4\qty(\braket{\partial_\theta\psi_\theta} - \abs{\braket{\psi_\theta}{\partial_\theta\psi_\theta}}^2).
		\end{equation}
	
		Similarly, the elements of SLD can be expressed in a closed form in terms of the eigenbasis of $\rho_\theta$ as
		
		\begin{equation}
			\frac{\hat{L}_{\rho_\theta,ij}}{2} = \frac{\delta_{ij}}{p_i + p_j} \pdv{p_i}{\theta} + \frac{p_j-p_i}{p_i + p_j} \braket{\psi_i}{\partial_\theta\psi_j},
		\end{equation}
		and the optimal measurement saturating the bound on
                the variance given by QFI can be given in terms of
                projectors $M_i = \op{l_i}$ on the eigenvectors
                $\ket{l_i}$ of the SLD. \rd{
                  If $\rho_\theta$ is a pure state, the SLD simplifes to $L_{\rho_\theta}\equiv L_\theta = \qty(\op{\partial_\theta \psi}{\psi} + \op{\psi}{\partial_\theta \psi})$ \cite{LYLW19}.}
	
		To illustrate the classical limit of the QFI, it is a rudimentary task to calculate this quantity for the rotation angle $\phi$,
%
		\begin{align}
			\qty(\bra{S,S}_{\sfrac{\pi}{2},\phi}\partial_\phi)\qty(\partial_\phi\ket{S,S}_{\sfrac{\pi}{2},\phi}) = S^2 + \frac{S}{2}, && 
			\abs{\qty(\bra{S,S}_{\sfrac{\pi}{2},\phi})\qty(\partial_\phi\ket{S,S}_{\sfrac{\pi}{2},\phi})}^2 = S^2.
		\end{align}
		This implies that QFI takes the form,
		\begin{equation}
			I_\phi(\ket{S,S}_{\sfrac{\pi}{2},\phi}) = 2S,
		\end{equation}
		which shows linear scaling with the number of the
                systems $N = 2S$, often referred to as the 
                ``classical limit'' or ``Standard Quantum Limit''
                (SQL), in agreement with the fact that and the state
                $\ket{S,S}_{\sfrac{\pi}{2},\phi}$ 
                is considered the most classical state of a spin
                system \cite{ACGT72,Giraud10}. 
		In the remainder of this paper we 
                demonstrate a protocol that allows quadratic scaling with $N$, thus the quantum counterpart of the classical limit. 
		
		
%
                	One can generalize the quantum Cram\'er-Rao bound to multi-parameter estimation. In such a case one considers the covariance matrix $\operatorname{Cov}(\boldsymbol{\theta}_{\text{est}})$ of an estimator $\boldsymbol{\theta}_{\text{est}}$, defined for a vector-valued parameter $\boldsymbol{\theta} := \qty{\theta_i}$. This leads to 
                  the bound \cite{H69,H11}
		\begin{equation} \label{eq:multiparam_est_bound}
			\operatorname{Cov}(\boldsymbol{\theta}_{\text{est}}) - \mathbf{I}_{\boldsymbol{\theta}}(\rho_{\boldsymbol{\theta}})^{-1} \geq 0\,,
		\end{equation}	
		where the quantum Fisher information matrix $\mathbf{I}_{\boldsymbol{\theta}}(\rho_{\boldsymbol{\theta}})$ is defined elementwise as
		\begin{equation}
			\qty(\mathbf{I}_{\boldsymbol{\theta}}(\rho_{\boldsymbol{\theta}}))_{ij} :=  \frac{1}{2}\Tr\left(\rho_{\boldsymbol{\theta}}\{L_{\theta_i},L_{\theta_j}\}\right)\,,
                      \end{equation}
                      where $\{\cdot,\cdot\}$ denotes the anti-commutator. 
In contrast to the single parameter quantum Cram\'er-Rao bound, the bound \eqref{eq:multiparam_est_bound} can usually not be saturated. A bound that can be up to a factor two tighter \cite{demkowicz-dobrzanski_multi-parameter_2020} and that can be saturated asymptotically in the limit of infinitely many samples was introduced by Holevo \cite{holevo_statistical_1973}, but it is, in general, more difficult to evaluate than the quantum Cram\'er-Rao bound. If the SLD operators for different parameters commute on average, $\tr \left(\rho_{\boldsymbol{\theta}}[L_{\theta_i},L_{\theta_j}]\right)=0$, then this Holevo-Cram\'er-Rao bound coincides with the quantum Cram\'er-Rao bound, such that also the latter can be saturated asymptotically \cite{RJD16,AFD19,tsang_quantum_2020,demkowicz-dobrzanski_multi-parameter_2020}. \rd{In particular, for pure states, $\rho_{\boldsymbol{\theta}} = \op{\psi}$, the average commutation relation is given by
\begin{equation} \label{eq:SLD_comm_purestates}
	\tr \left(\rho_{\boldsymbol{\theta}}\comm{L_{\theta_i}}{L_{\theta_j}}\right) = 8i\Im\qty(\bra{\partial_{\theta_i}\psi}\qty(\mathbbm{1} - \op{\psi})\ket{\partial_{\theta_j}\psi})
\end{equation}
}
	\section{Results}

	\subsection{Selected axes for single-spin maximal squeezing}
	
		Following the steps of SU(1,1) interferometry in the
                formulation by Caves, we translate the
                squeeze-\textbf{displace}-unsqueeze procedure into its spin
                equivalent, squeeze-\textbf{rotate}-unsqueeze (SRU),
                closely related to the already successful experimental
                delta-kick squeezing (DKS) protocol \cite{CGSP21}. We
                begin the extension of the interferometric scheme to
                spins by considering a toy example of a single
                spin. First we define the SRU protocol acting on a SCS
                along the $X$ axis, which serves as an equivalent for
                the bosonic vacuum
              $\ket{0}$,
		\begin{align} 
			\ket{\text{SRU}_1(\gamma,\Phi;\mu)} = \underbrace{e^{-i \mu S_z^2}}_{\text{Unsqueeze}} \underbrace{e^{i \gamma S(\Phi)}}_{\text{Rotate}} \underbrace{e^{i \mu S_z^2}}_{\text{Squeeze}} \underbrace{\ket{S, S}_{\pi/2,0}}_{\text{SCS along X axis}} \label{eq:1_spin_SRU}
		\end{align}
		with $S(\Phi) = \cos\Phi S_x + \sin\Phi S_y$. In particular, such a rotation could be induced by interaction with the magnetic field via standard Hamiltonian $H = -\boldsymbol{\mu}\cdot\bf{B}$. with the magnetic moment proportional to the spin vector, $\boldsymbol{\mu} \propto \bf{S}$. In such case the angle $\gamma$ could be used to estimate either the time of interaction $t$ for constant field $\bf{B}$ or the magnitude of field $\norm{\bf{B}}$ for a fixed direction of field and interaction time. Similar unitary operators were used in investigation of the phenomenon of quantum chaos in the model of the quantized kicked top \cite{WBY85, KSH87, HS88, BGHS96}. As a further simplification we limit the axis around which the spin can rotate to $\Phi \in\qty{0,\sfrac{\pi}{2}}$. If we consider the case of the maximal squeezing, $\mu = \sfrac{\pi}{2}$ (as illustrated in the rightmost panel of Fig.~\ref{fig:wignershape}), we find that the states corresponding to the SRU protocol with rotations around $X$ and $Y$ axes are particularly simple. Depending on the spin being full- or half-integer the axes interchange and subtle details of the states are affected, as listed below,
		
		\begin{enumerate}
			\item[a)] Integer spin,
			\begin{align}
				\ket{\text{SRU}_1(\gamma,\sfrac{\pi}{2};\sfrac{\pi}{2})} = & \frac{1}{2}\left(\ket{S,S}_{\sfrac{\pi}{2} - \gamma,0} - (-1)^S i\ket{S,S}_{\sfrac{\pi}{2}+ \gamma,\pi} \right. \nonumber\\
					& \left. + \ket{S,S}_{\sfrac{\pi}{2} + \gamma,0} + (-1)^S i \ket{S,S}_{\sfrac{\pi}{2} - \gamma,\pi} \right), \label{eq:intSpinPi2}\\
				\ket{\text{SRU}_1(\gamma,0;\sfrac{\pi}{2})} = & \cos(S\gamma) \ket{S,S}_{\pi/2,0} - (-1)^S\sin(S\gamma )\ket{S,S}_{\pi/2,\pi},  \label{eq:intSpin0}
			\end{align}
			\item[b)] Half-integer spin,
			\begin{align}
				\ket{\text{SRU}_1(\gamma,0;\sfrac{\pi}{2})} = & \frac{1}{2}\left(\ket{S,S}_{\sfrac{\pi}{2} + \gamma,0} - e^{-i\pi S} \ket{S,S}_{\sfrac{\pi}{2}- \gamma,\pi} \right. \nonumber\\
				&	\left. + \ket{S,S}_{\sfrac{\pi}{2} - \gamma,0} - e^{i\pi S} \ket{S,S}_{\sfrac{\pi}{2} + \gamma,\pi} \right),\\
				\ket{\text{SRU}_1(\gamma,\sfrac{\pi}{2};\sfrac{\pi}{2})} = & \cos(\gamma S) \ket{S,S}_{\sfrac{\pi}{2},0} + (-1)^{S-\frac{1}{2}} \sin(\gamma S) \ket{S,S}_{\sfrac{\pi}{2},\pi}.
			\end{align}
		\end{enumerate}
		
		The Fisher information for the rotation angle $\gamma$ around the two selected axes are given by the following expression,
		
	
		\begin{equation}
			I\qty(\ket{\text{SRU}_1(\gamma,\Phi;\sfrac{\pi}{2})};\gamma) = 
			\begin{tabular}{c | c c}
				& $\Phi = 0$ & $\Phi = \sfrac{\pi}{2}$ \\ \hline
				\text{integer spin} & $4S^2 = N^2$ & $2S = N$ \\
				\text{half-integer spin} & $2S = N$ & $4S^2 = N^2$
			\end{tabular},
		\end{equation}
		where $N = 2S$ is the number of spins-$1/2$ in a composite system setting. This illustrates that for both chosen axes ($\Phi = 0$ or $\pi/2$) we find both characteristic scalings of the QFI, depending on whether the spin is integer or half-integer. {Note that this is in alignment with  already known results from the bosonic SDU protocol -- one direction is found to have high QFI, while the other remains low.}
		
		Focusing first on the integer spin, note that for the $\Phi = \pi/2$ case in \eqref{eq:intSpinPi2}, which does not represent an advantage for rotation estimation over the normal SCS, the aspect of the state that varies with $\gamma$ is the polar angle of all four components. In the $\Phi = 0$ case in \eqref{eq:intSpin0}, which indeed yields the QFI scaling with the square of the number of spins $1/2$, the coefficients in front of two constant and orthogonal SCS vary with the rescaled angle of rotation, $S\gamma$. Thus, the SDU protocol may be seen as an operation that amplifies the rotation $\gamma$ by a factor $S$ and transfers it to the 2-dimensional subspace spanned by highest and lowest-weight states of the representation. Analogous considerations apply to the half-spin with small adjustments needed in the phases of the components and axes $X$ and $Y$ interchanged. Detailed calculations are given in Appendix \ref{app:1spin_selected_ax}.
	
		The optimal measurement, which saturates the Heisenberg limit, can be exemplified by considering the SLD for the state $\ket{\text{SRU}_1(\gamma,0;\sfrac{\pi}{2})}$ for integer spin. In this case the non-zero components of SLD can be expressed in terms of two spanning states
		\begin{align}
			\ket{\psi_1} & = \ket{\text{SRU}_1(\gamma,0;\sfrac{\pi}{2})} \propto \partial_\gamma \ket{\psi_2}, \\
			\ket{\psi_2} & = \sin(\gamma S) \ket{S,S}_{\sfrac{\pi}{2},0} + \cos(\gamma S) \ket{S,S}_{\sfrac{\pi}{2},\pi} \propto \partial_\gamma \ket{\psi_1}, 
		\end{align}
		which in this particular case give $L_{\ket{\text{SRU}_1(\gamma,0;\sfrac{\pi}{2})}} = i\gamma S \sigma_y$. Eigenvectors of this subspace are easily calculated to be
		
		\begin{equation}
			\ket{\psi_\pm} = \frac{1}{\sqrt{2}} \qty(\ket{\psi_1} \pm i \ket{\psi_2}) \propto \frac{1}{\sqrt{2}}\qty(\ket{S,S}_{\sfrac{\pi}{2},0} + i\ket{S,S}_{\sfrac{\pi}{2},\pi}).
		\end{equation}
		This form has an additional advantage of being independent from the angle of rotation, allowing for a non-adaptive measurement scheme. 
		
		Observe that a squeezed state with squeezing parameter $\mu = \sfrac{\pi}{2}$ would be considered severely oversqueezed by the standards set in \cite{KU93}, as it does not minimize the uncertainty in any of the axes orthogonal to the original axes of the SCS. However, we highlight here that $\ket{S;\sfrac{\pi}{2}}_{\sfrac{\pi}{2},0}$ is in fact a maximally entangled state, superposition of the highest and lowest weight states, providing an intuitive explanation of reaching the Heisenberg limit. 

	\subsection{QFI for arbitrary equatorial axis for the two-spin scenario}
	
		After demonstrating explicitly that the single-system procedure allows us to reach the QFI scaling with the square of the number of spins $1/2$ we now shift our attention to a scenario involving the main spin $S_M$ and a probe spin $S_P$. The SRU protocol for such a composite system can be defined in analogy to the single spin case as
		
		\begin{equation} \label{eq:SRU_2spin_prot}
			\ket{\text{SRU}_2(\gamma,\Phi;\mu)} = \underbrace{e^{-i \mu S_z\otimes S_z}}_{\text{Unsqueeze}} \underbrace{e^{i \gamma S(\Phi)\otimes\mathbb{I}}}_{\text{Rotate}} \underbrace{e^{i \mu S_z\otimes S_z}}_{\text{Squeeze}} \underbrace{\ket{S_M, S_M}_{\pi/2,0}\otimes\ket{S_P, S_P}_{\pi/2,0}}_{\text{Two SCS along X axis}}.
		\end{equation}
		The squeeze and unsqueeze steps in this case are performed using the two-spin single-axis squeezing Hamiltonian, and the rotation is assumed to be applied only to the main system. 
		
		\begin{figure}[H]
			\centering
			\includegraphics[width=.67\linewidth]{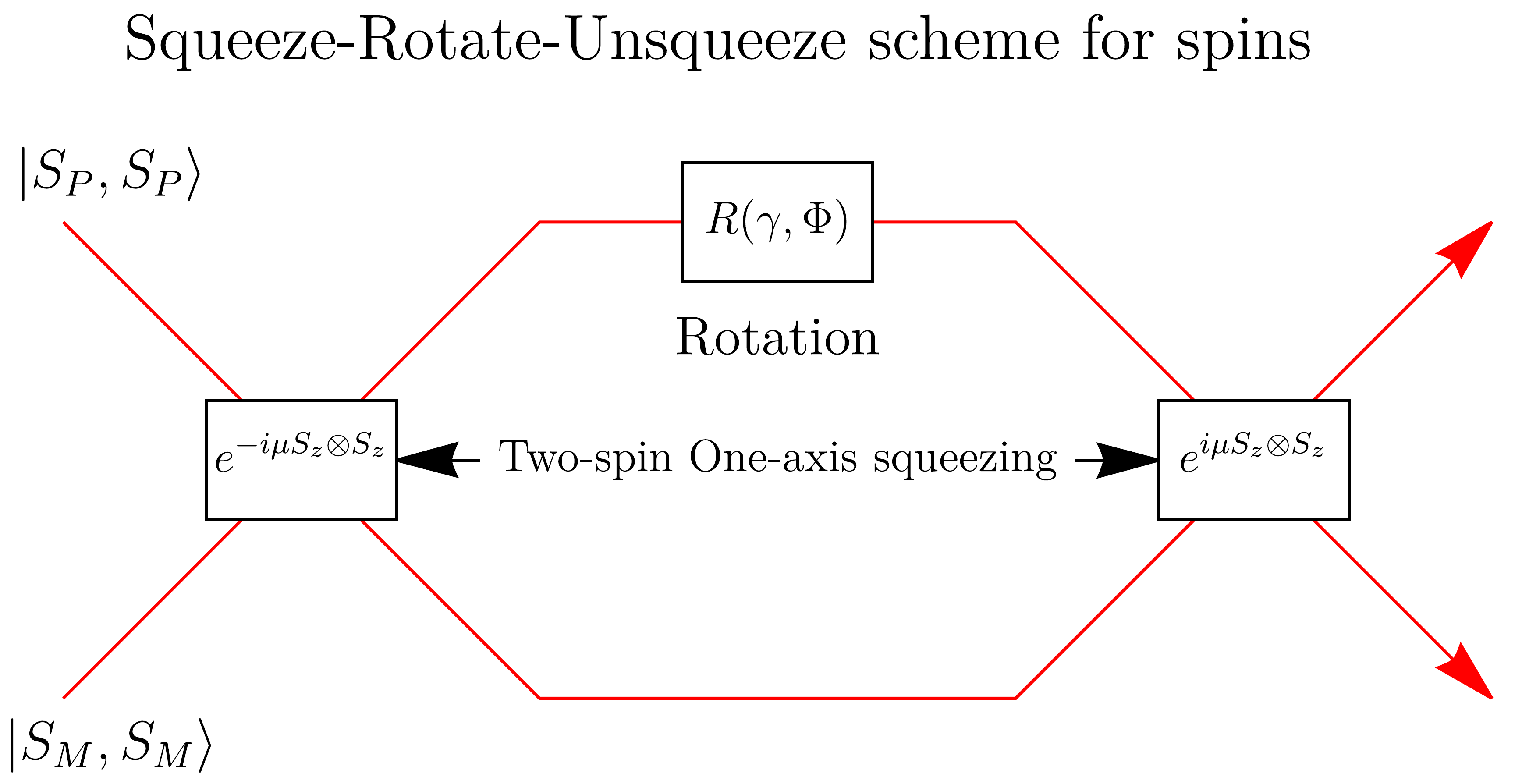} 
			\caption{
				Schematic depiction of the SRU protocol described by equation \eqref{eq:SRU_2spin_prot} with a rotation $R(\gamma,\Phi)$ only on one of the spins and the two-spin squeeze operators $\exp(i\mu S_z\otimes S_z)$. Note the similarity between the introduced SRU scheme and the $SU(1,1)$ interferometer.}
			\label{fig:Spin2_interferometer}
		\end{figure}
	
		After rudimentary calculations one arrives at the explicit expression for the state after the SRU protocol,
		
		\begin{equation}
			{\footnotesize
			\begin{alignedat}{3}
				&\ket{\text{SRU}_2(\gamma,\Phi;\mu)}& = \frac{1}{2^{S_M + S_P}}\sum_{j=-S_M}^{S_M} \sum_{k = -S_P}^{S_P}
				\sqrt{\binom{2S_M}{S_M - j}\binom{2S_P}{S_P - k}}
				 &\qty(\cos\frac{\gamma}{2} + e^{i(\Phi-k \mu)}\sin\frac{\gamma}{2})^{S_M + j} \nonumber& \\
				&&\times& \qty(\cos\frac{\gamma}{2} - e^{-i(\Phi-k \mu)}\sin\frac{\gamma}{2})^{S_M - j} \ket{j,\,k},&
			\end{alignedat}}
		\end{equation}
		where we use the shorthand notation $\ket{j,k} \equiv \ket{S_M,j}\otimes\ket{S_P,k}$. The above form allows us to derive the expression for the QFI for the rotation angle $\gamma$ around any fixed equatorial axis parametrized by the angle $\Phi$,
		
		\begin{equation}
			{\footnotesize
			I_\gamma(\ket{\text{SRU}_2(\gamma,\Phi;\mu)}) = S_M \qty(2 S_M+1 - \qty(2 S_M - 1) \cos (2 \Phi ) \cos ^{2 S_P}(\mu )-4 S_M \sin ^2(\Phi ) \cos ^{4 S_P}\qty(\frac{\mu }{2})).} \label{eq:spin_QFI_anyAxis}
		\end{equation}
		See details in Appendix \ref{app:2spin_equat_axes}. In Fig.~\ref{fig:QFI_Plots} we present some properties of the above function for a fixed value of $S_M = 5$ for different values of $S_P$ and $\Phi$. If the squeezing $\mu = \pi$, QFI reaches the value $I = 4S_M^2 = 100$ for all probe spins $S_P$ for either $X$ or $Y$ axis, depending on half- or integer-spin of the probe. At the same time the estimation for the orthogonal axis yields QFI at the level of $I = 2 S_M = 10$. As $S_P$ grows, the plateau for $\Phi = \pi/4$ at $I = (4S^2 + 2S)/2 =  55$ around $\mu = \pi/2$ grows and flattens.

		\begin{figure}[H]
			\centering
			\includegraphics[width=.75\linewidth]{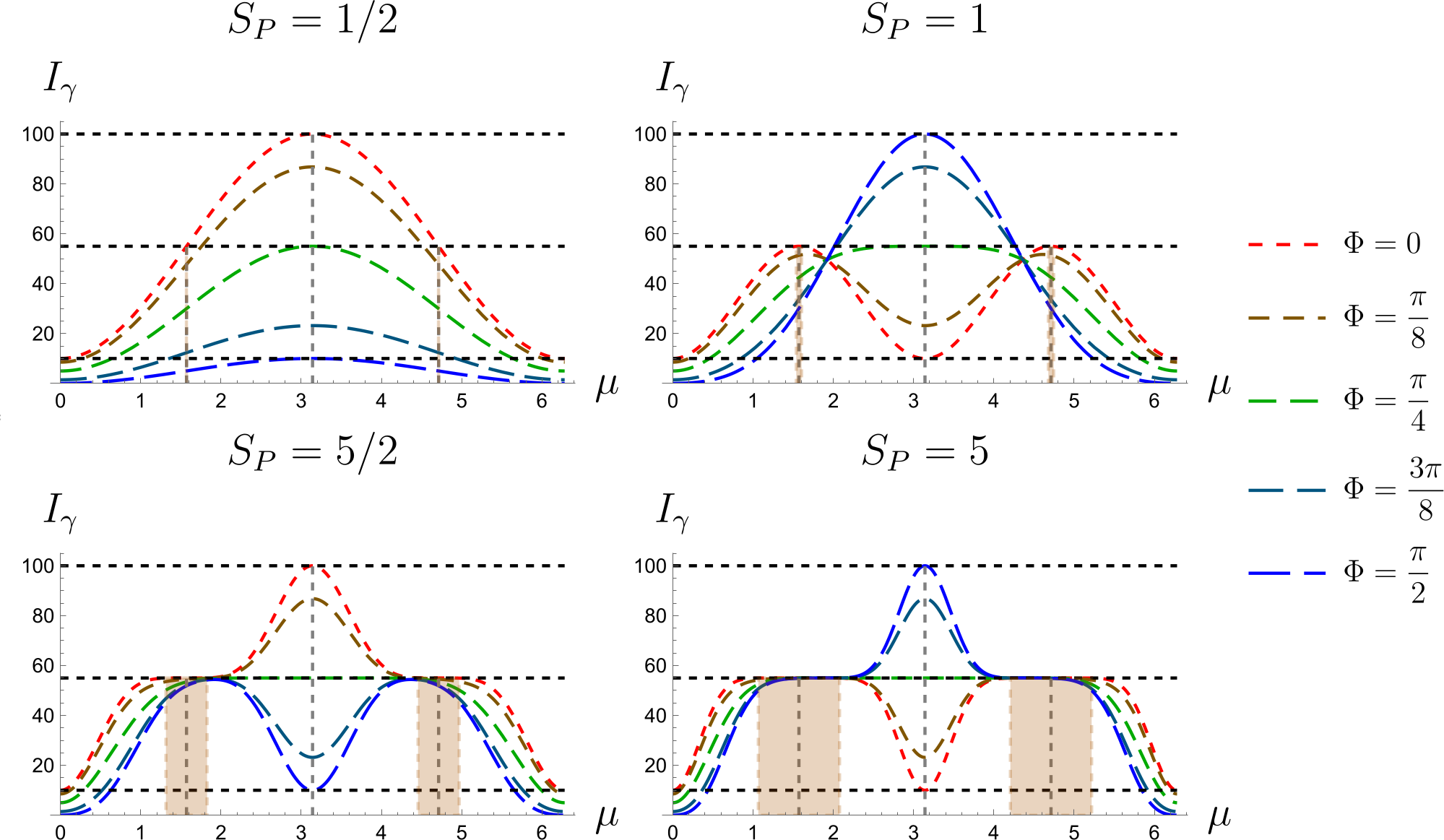} 
			\caption{
				Fisher information $I_\gamma$ for the SRU protocol as a function of the squeezing parameter $\mu$ and rotation axis angle $\Phi$ (Eq. \eqref{eq:spin_QFI_anyAxis}) for a fixed main spin, $S_M = 5$. Note the maximal value for $\mu = \pi$ and the plateau around $\pi/2$ (shaded region), growing with the probe spin $S_P$.}
			\label{fig:QFI_Plots}
		\end{figure}
	
		The qualitative features that can be seen from the Fig.~\ref{fig:QFI_Plots} will be now formalised, with special attention for maximal squeezing $\mu = \pi$ and the half-maximal-squeezing with $\mu = \sfrac{\pi}{2}$.
	
		The former case is exceedingly simple and reduces to a formula reminiscent of a sum of coordinates of an ellipse,
		\begin{align}
			I_\gamma(\ket{\text{SRU}_2(\gamma,\Phi;\pi)}) & = S_M \qty(2 S_M+1 + \qty(1-2 S_M) \cos (2 \Phi ) (-1)^{2 S_P}) \\
			& = 2S_M \sin[2](\Phi + S_P \pi) + 4S_M^2 \cos[2](\Phi + S_P \pi),
		\end{align}
		bounded from above and below by
		\begin{equation}
			2S_M \leq I_\gamma\left(\ket{\text{SRU}_2(\gamma,\Phi;\pi)}\right)  \leq 4S_M^2,
		\end{equation}
		where the axes for which the bounds are saturated depend solely on $S_P$. More precisely, $X$ axis is optimal for half-spins, whereas $Y$ axis reaches optimality for integer spins. This aligns neatly with the properties found for the same axes for one-spin case.
		
		In the less obvious case of the half-maximal squeezing, we find that around $\mu = \frac{\pi}{2}$ the QFI achieves a plateau at a level between the linear and quadratic scaling with the number of spins $1/2$. The observation is made rigorous by considering an approximation of the QFI with $\mu = \sfrac{\pi}{2} + \epsilon$ and $\epsilon \ll 1$,
		
		\begin{align}
			I_\gamma(\ket{\text{SRU}_2(\gamma,\Phi;\sfrac{\pi}{2})})
			\overset{(a1)}{\approx} & \;S_M \left[2 S_M+1 + \qty(1-2 S_M) \cos (2 \Phi ) \epsilon^{2S_P}\qty(1 + \mathcal{O}\qty(\epsilon^{2}))\right. \nonumber \\
			& \;\left.-4 S_M \sin ^2(\Phi )2^{-2S_P}\qty(1-2 S_P \epsilon + \mathcal{O}\qty(\epsilon^{2}))\right] \\
			\overset{(a2)}{\approx} & \;S_M \qty(2 S_M+1 + \qty(1-2 S_M) \cos (2 \Phi ) \epsilon^{2S_P}) \\
			\overset{(a3)}{\approx} & \;\frac{4S_M^2 + 2S_M}{2}, \label{eq:2S1Ax_QFI_piHalf_semiOptimal}
		\end{align}
		with the approximation $(a1)$ achieved by Taylor
                expansion around $\epsilon = 0$ in the lowest order,
                $(a2)$ by noting that $2^{-2 S_P}$ vanishes in the
                limit of large $S_P$, and $(a3)$ by noting that
                $\epsilon\ll 1$ justifies neglecting the term of order
                $\epsilon^{2S_P}$. This provides the formal
                demonstration that the QFI is flat around $\mu =
                \sfrac{\pi}{2}$ and becomes progressively more flat
                for large probe spins.  The final line demonstrates
                that we are exactly at the ``optimal'' value halfway
                between the Heisenberg and the classical scalings,
                giving a stable region within which rotation around
                any axis can be estimated. In particular, in the
                stable region, one still retains
                $I_\gamma\qty(\ket{\text{SRU}_2(\gamma,\Phi;\sfrac{\pi}{2})})
                \simeq O(N_M^2)$ for $S_M = N_M/2$. \rd{Similar QFI
                  values for the estimation of the rotation angle 
                  with fixed rotation axis using states at
                  half-maximal squeezing have been recently reported
                  in \cite{Metal23}. Note that the independence of QFI from the choice of the rotation exhibited by the SRU procedure
                  mimics the behaviour of QFI for SU(1,1)
                  interferometric protocol, where any direction of
                  displacement has the same high value of QFI,
                  allowing for estimation with enhanced sensitivity in
                  arbitrary direction.
As such the results we demonstrate for arbitrary equatorial axis can be seen as an extension of results from \cite{Metal23}. 
}
		
		It is instructive to compare the expression \eqref{eq:spin_QFI_anyAxis} with a general formula for the QFI for a single spin,
		\begin{equation} \label{eq:QFI_1spin}
		{\small
			I_\gamma\qty(\ket{\text{SRU}_1(\gamma,\Phi;\mu)}) = S \qty(2 S+1 - (2 S-1) \cos (2 \Phi ) \cos ^{2 S-2}(2 \mu )-4S \cos ^2(\Phi ) \cos ^{4 S-2}(\mu ))
		}
		\end{equation}
		which is almost identical except for the replacement $\mu \rightarrow 2\mu$ and some minor changes in the powers and trigonometric functions of the angle $\Phi$. The full derivation of this expression is provided in Appendix \ref{app:1spin_equat_axes}. Most importantly, the analysis concerning the plateau $\ket{\text{SRU}_2(\gamma,\Phi;\sfrac{\pi}{2})}$ applies without modification for $\ket{\text{SRU}_1(\gamma,\Phi;\sfrac{\pi}{4})}$. This suggests that the state $\ket{S;\sfrac{\pi}{4}}$ presented in the middle panel of Fig. \ref{fig:wignershape}, similar in the qualitative properties of its Wigner function to the compass state \cite{Z01}, may be of independent interest. 
		
		\begin{figure}[H]
			\centering
			\includegraphics[width=.75\linewidth]{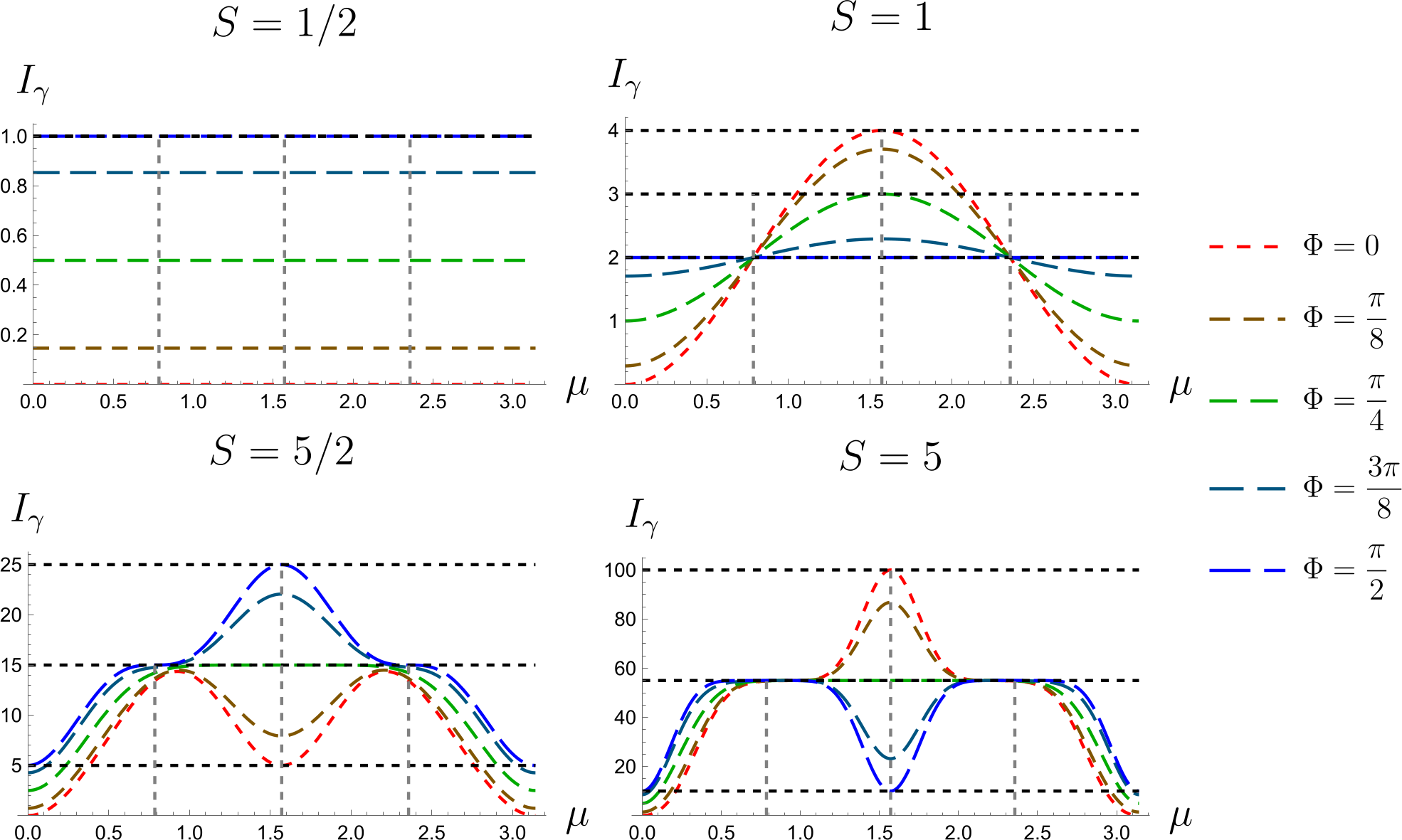} 
			\caption{
				Single-spin counterpart of Fig. \ref{fig:QFI_Plots} for the same range of spin values. Note the reduced range of the squeezing parameter $\mu$, from $\qty[0,2\pi]$ to $\qty[0,\pi]$. The plateau around $\mu = \sfrac{\pi}{4}$ flattens out with growing $S$, in analogy with the effect of enlarging the probe spin $S_P$ in the two-spin setting.}
			\label{fig:QFI_Plots_1spin}
		\end{figure}
	
	\subsection{Different ways of \rd{en}coding a relative rotation}
        	In \cite{JD12} it was pointed out that different ways of coding a relative phase between the two modes in an interferometer can lead to different values of the QFI and hence the sensitivity: if the first arm acquires a phase $\varphi/2$ and the second a phase $-\varphi/2$, the minimal error for estimating $\varphi$ was shown to be substantially smaller than if the first arm carries $\varphi$ and the second the phase 0.  From the classical perspective of a Mach-Zehnder interferometer one would expect that only the relative phase between the two arms matters, because that is all the resulting interference pattern in the two output ports of the interferometer depends on. Quantum-mechanically, however, the two states created are different and can hence lead to different QFI. This implies that there must be measurement schemes that provide higher sensitivity than the usual closing of the interferometer and detection of intensities in the output ports. \\ 

          The analogous situation was later studied in the context of SU(1,1) interferometry in \cite{YEtAl19}.
          More generally, the second mode or spin can be considered a quantum reference frame \cite{aharonov_quantum_1984,bartlett_reference_2007,giacomini_quantum_2022} that need not be in a classical state and can even be entangled with the main system, i.e.~the first mode or spin.
          Here we investigate a similar question that may arise in the proposed SRU protocol. We consider a rotation $R(\gamma_1, \Phi_1)$ on the main spin and an additional rotation $R(\gamma_2, \Phi_2)$ on the probe spin. Within this context we consider the multiparameter estimation problem for $\boldsymbol{\gamma} = \qty(\gamma_1,\gamma_2)$.
		\begin{figure}[H]
			\centering
			\includegraphics[width=.75\linewidth]{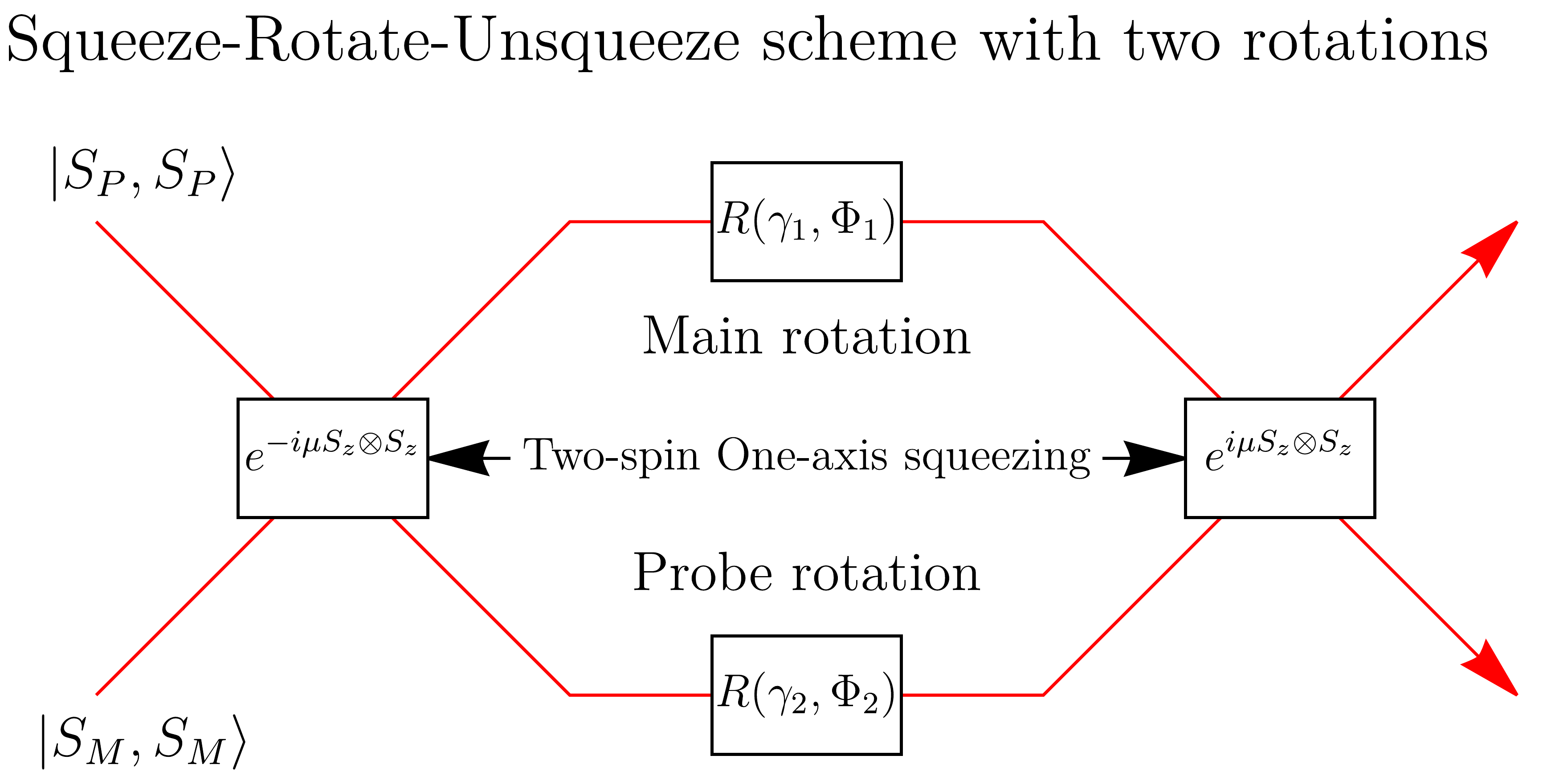} 
			\caption{
					The additional rotation $R(\gamma_2, \Phi_2)$ on the probe spin is introduced in order 
to investigate how the coding of the rotation angle relative to the probe spin that serves as reference influences the sensitivity, in analogy to the work for the SU(1,1) interferometers \cite{YEtAl19}.} 
			\label{fig:scheme_Spin2_2_rots}
		\end{figure}
	
			The quantum Fisher information matrix reads
		\begin{equation}
			I_{\boldsymbol{\gamma}} = \mqty(I_{\gamma_1} & \qty(I_{\boldsymbol{\gamma}})_{12} \\ \qty(I_{\boldsymbol{\gamma}})_{21} & I_{\gamma_2}),
		\end{equation}
		where for convenience we dropped the arguments. The diagonal elements are given by the already known expressions for the main spin
		{\small
		\begin{equation}
			\begin{aligned}
			I_{\gamma_1} & \equiv I_{\gamma_1}\qty(\ket{\text{SRU}_2(\gamma_1,\gamma_2,\Phi_1,\Phi_2;\mu)}) \\
			& = S_M \qty(2 S_M+1 - \qty(2 S_M - 1) \cos (2 \Phi_1 ) \cos ^{2 S_P}(\mu )-4 S_M \sin ^2(\Phi_1 ) \cos ^{4 S_P}\qty(\frac{\mu }{2})),
		\end{aligned}
		\end{equation}}
		and similarly for $\gamma_2$, with the roles of $S_M, \Phi_1$ and $S_P, \Phi_2$ swapped,
		{\small
			\begin{equation}
				\begin{aligned}
					I_{\gamma_2} & \equiv I_{\gamma_2}\qty(\ket{\text{SRU}_2(\gamma_1,\gamma_2,\Phi_1,\Phi_2;\mu)}) \\
					& = S_P \qty(2 S_P+1 - \qty(2 S_P - 1) \cos (2 \Phi_2 ) \cos ^{2 S_M}(\mu )-4 S_P \sin ^2(\Phi_2 ) \cos ^{4 S_M}\qty(\frac{\mu }{2}))\,.
				\end{aligned}
	\end{equation}}
	The off-diagonal terms can be found as the covariance between $\gamma_1$ and $\gamma_2$,
	{\small
		\begin{equation}
			\begin{aligned}
				\qty(\mathbf{I}_{\boldsymbol{\gamma}})_{12} = \qty(\mathbf{I}_{\boldsymbol{\gamma}})_{21} & = 4\qty(\ev{S(\Phi_1)\otimes S(\Phi_2)} - \ev{S(\Phi_1)}\ev{S(\Phi_2)}) \\
				& = 4 S_P S_M \cos\Phi_1\cos\Phi_2 \cos[2(S_P + S_M - 1)](\frac{\mu}{2})\sin[2](\frac{\mu}{2})\,.
			\end{aligned}
	\end{equation}}
	From these the lower bounds on the variance of the estimators of $\gamma_1$ and $\gamma_2$ angles, calculated from positivity of the primary minors of $\operatorname{Cov}(\boldsymbol{\gamma}) - \mathbf{I}_{\boldsymbol{\gamma}}$, are given as
	\begin{align}\label{eq:multiparam_QFI}
		\operatorname{Var}(\gamma_1) & \geq \frac{I_{\gamma_2}}{\det(\mathbf{I}_{\boldsymbol{\gamma}})}, &
		\operatorname{Var}(\gamma_2) & \geq \frac{I_{\gamma_1}}{\det(\mathbf{I}_{\boldsymbol{\gamma}})}.
	\end{align}

	
	\begin{figure}[H]
		\centering
		\includegraphics[width=\linewidth]{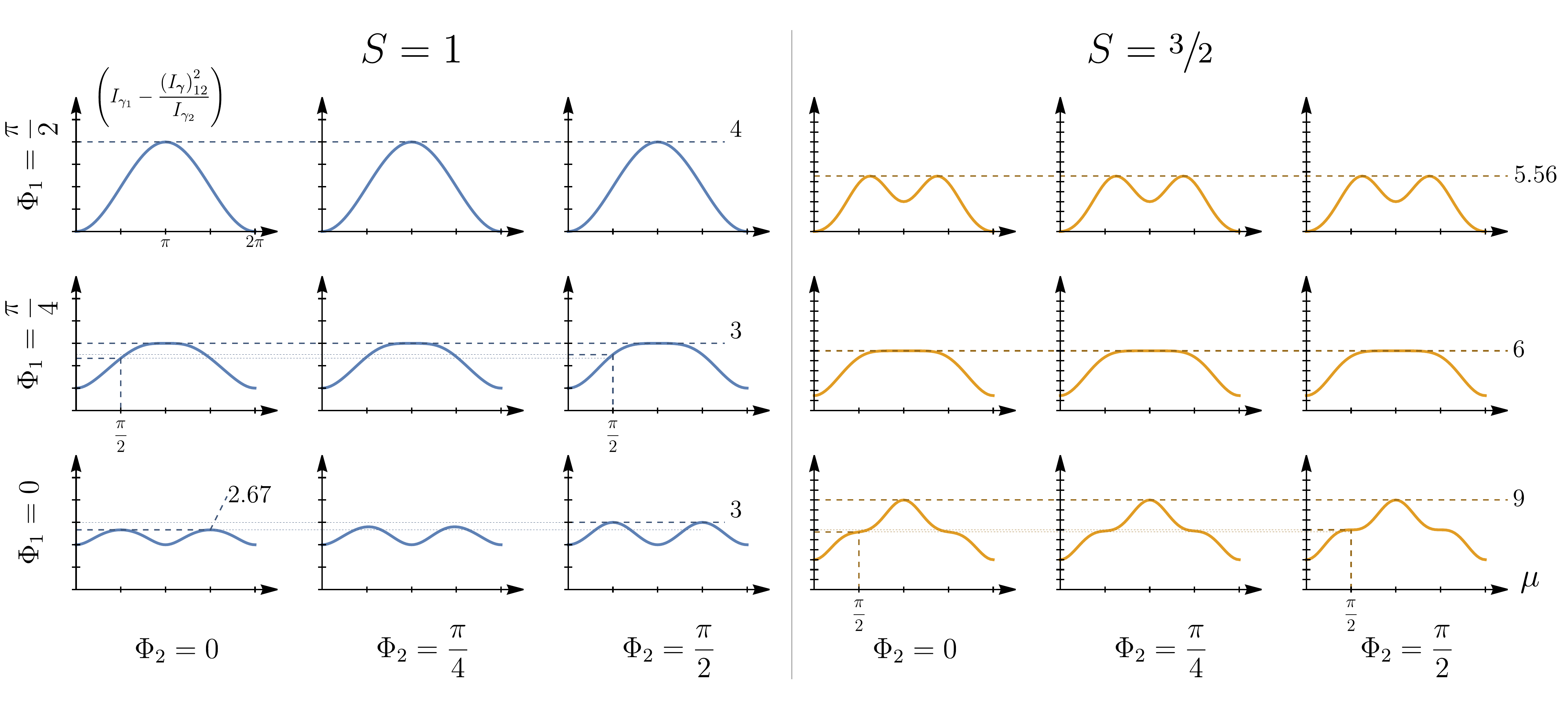} 
		\caption{Bounds on the sensitivity for estimating $\gamma_1$ in the scheme as shown in Fig. \ref{fig:scheme_Spin2_2_rots} as function of the squeezing parameter $\mu$, for both main and probe spins equal and for selected angles $\Phi_1$ and $\Phi_2$ of rotation axes. Left panel: 
                  $S=1$ 
                  right panel: $S=3/2$.  Combinations of the same relative rotation angle $\Phi_2-\Phi_1$ are on the lines perpendicular to the diagonal of the matrix of plots and seen to give quite different lower bounds $(I_{\gamma_1} - \frac{\qty(I_{\boldsymbol{\gamma}})_{12}^2}{I_{\gamma_2}})^{-1} $ for the error of the $\gamma_1$ estimation. Note that the optimal bounds at $\mu = \pi/2$ are achieved at $\Phi_1 = \pi/2$ and $\Phi_1 = 0$ for $S=1$ and $3/2$ respectively. Furthermore, for $S = 1$ and $\Phi_1 = 0$ we note decrease of the bound from $3$ to $2.67$ together with decreasing $\Phi_2$ from $\pi/2$ to $0$, marked by thin lines.%
                }
		\label{fig:multiparameter}
	\end{figure}

	We note that $\qty(I_{\boldsymbol{\gamma}})_{12}\propto \cos\Phi_1\cos\Phi_2$ and therefore can be set to zero by setting $\Phi_1 = \pi/2$ or $\Phi_2 = \pi/2$. 
	Next, we may focus on the denominator of the first of the expressions \eqref{eq:multiparam_QFI} and its qualitative properties by studying the case of $S_P = S_M = S$ for values $S = 1$ and $S = \sfrac{3}{2}$ (see Fig.~\ref{fig:multiparameter}). For $S=1$ and $\Phi_1 = 0$ we can see clear dependence of the maximal value on the angle $\Phi_2$, with minimum of $8/3 \approx 2.67$ and maximum of $3$, coinciding with already known maximum in the single-parameter estimation in similar setting. For $\Phi_1 = \frac{\pi}{4}$ we see the already known central maximum with the value $3$. However, a small change can be seen at half-squeezing, $\mu = \pi/2$. 
        It follows from the fact that the correction term $\frac{\qty(I_{\boldsymbol{\gamma}})_{12}^2}{I_{\gamma_2}}$ is propotional to $\sin[2](\mu)$, which vanishes for coherent and maximally mixed states, $\mu = 0$ and $\mu = \pi$, respectively. A similar dependence can be seen for $S = \sfrac{3}{2}$. The effect, however, is diminished as the corrections are linear in both main and auxiliary spin, thus not significant in comparison to quadratic terms.
        Furthermore, they depend on $\sin[2](\frac{\mu}{2})\cos[2(S_M + S_P -1)](\frac{\mu}{2})$. For $S_M = S_P = S$ we have the maxima of this expression attained at $\mu_{\max} = 4 \arctan\left(\sqrt{2 S-1}\pm\sqrt{2 S}\right)$ with $\sin[2](\frac{\mu_{\max}}{2})\cos[4S - 2](\frac{\mu_{\max}}{2}) = -\frac{(1-2S)^{2S-1}}{(4S)^S} \approx O(S \log S)$, smaller than the leading order of $S^2$, thus negligible for $2S \gg 1$.

		

	In order to draw a proper analogy between a phase difference of the bosonic interferometry, considered in \cite{JD12}, and the rotation angle difference in the introduced SRU scheme, one has to note that such a difference is not well defined, unless we set $\Phi_1 = \Phi_2 := \Phi$. In such a case one can consider sensitivity bounds for the relative rotation, $\Delta\boldsymbol{\gamma} = \gamma_2 - \gamma_1$. If we assume that the estimators for both rotations are uncorrelated, $\operatorname{Cov}(\gamma_1,\,\gamma_2) = 0$, we may proceed to find
	
	\begin{equation} \label{eq:rotation_dif_1}
		\operatorname{Var}(\gamma_2 - \gamma_1) = 
		\operatorname{Var}(\gamma_2) + \operatorname{Var}(\gamma_1) \geq 
		\frac{I_{\gamma_1} + I_{\gamma_2}}{\det(\mathbf{I}_{\boldsymbol{\gamma}})} \equiv 
		\tilde{I}_{\Delta \boldsymbol{\gamma}}^{-1}.
	\end{equation}
	
	\begin{figure}[H]
		\centering
		\includegraphics[width=\linewidth]{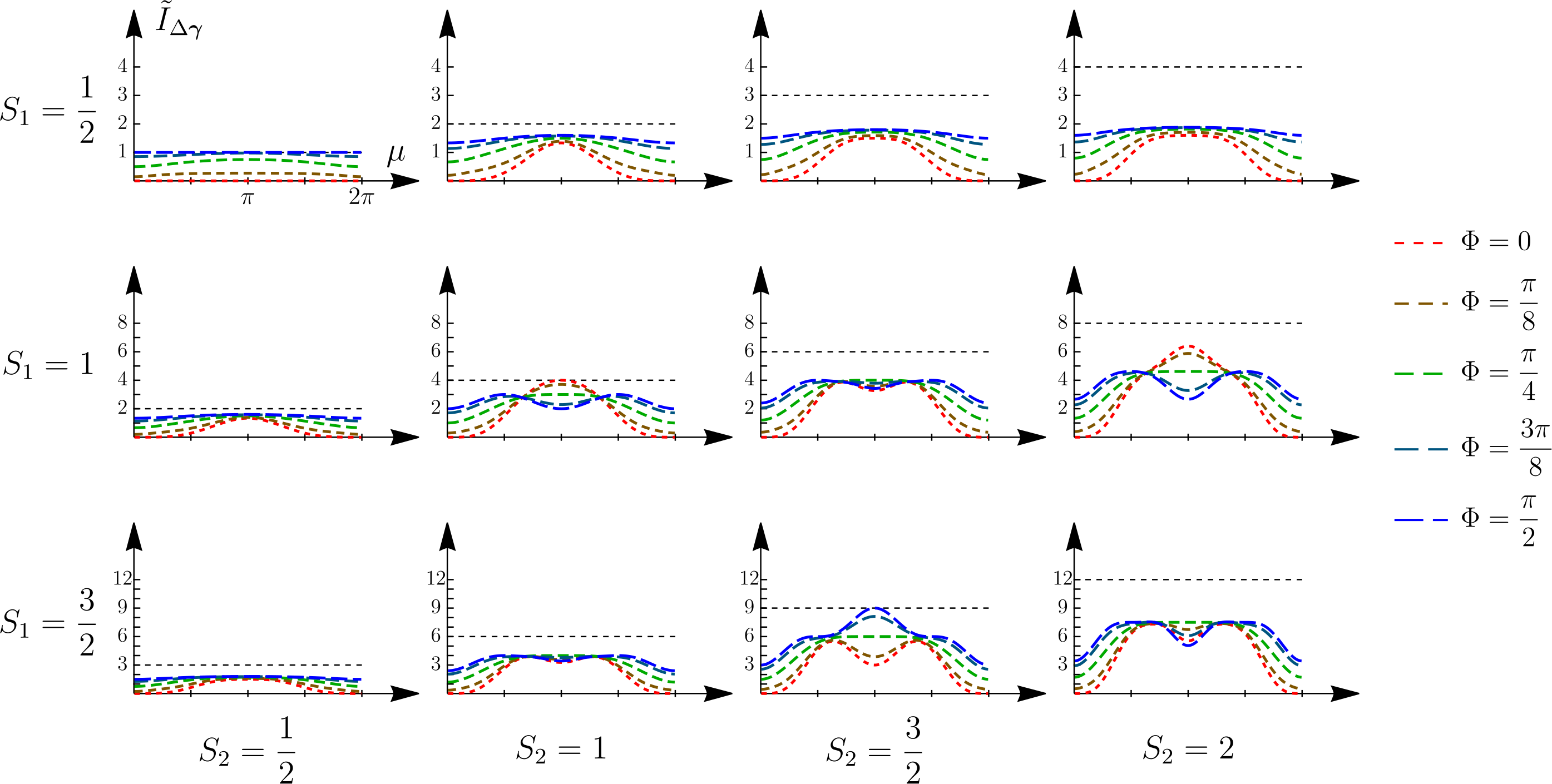}
		\caption{Plots of values of $\tilde{I}_{\Delta \boldsymbol{\gamma}}$ defined in \eqref{eq:rotation_dif_1} for selected values of $S_1,\,S_2$ and $\Phi$. Note the dashed black line, corresponding to the value $\tilde{I}_{\Delta \boldsymbol{\gamma}} = 4S_1S_2$, which is saturated only when $S_1 = S_2 = S$, $\mu = \pi$ and the axis of rotation is either $X$ or $Y$, depending whether $S$ is integer or half-integer.}
		\label{fig:qfirotationdifference}
	\end{figure}
	
	The following expression is depicted in Fig. \ref{fig:qfirotationdifference}. Note that for equal spins, $S_1 = S_2 = S$, the bound $\tilde{I}_{\Delta \boldsymbol{\gamma}}$ behaves similarly to the Fisher information $I_\gamma(\ket{\text{SRU}_2(\gamma,\Phi;\pi)})$ as depicted in Fig. \ref{fig:QFI_Plots}. In particular, we still see the plateau region around the value $2S^2 + S$ for half-squeezing $\mu = \frac{\pi}{2}$ for any axis angle $\Phi$. Similar plateau regions persist for any combination of $S_1 \neq S_2$. Finally, we note that for $S_2 - S_1$ being half-integer the central maximum is replaced by an indent for both $\Phi = 0$ and $\Phi = \frac{\pi}{2}$.

	On the other hand, if we allow covariance to be non-zero, by considering the semipositivity condition \eqref{eq:multiparam_est_bound} we find the bounds on covariance given in a compact way by	
	\begin{equation}
	\abs{\operatorname{Cov}(\gamma_1, \gamma_2) + \frac{(\mathbf{I}_{\boldsymbol{\gamma}})_{12}}{\det(\mathbf{I}_{\boldsymbol{\gamma}})}} \leq 
	\sqrt{
		\qty(\operatorname{Var}(\gamma_1) - \frac{I_{\gamma_2}}{\det(\mathbf{I}_{\boldsymbol{\gamma}})})
		\qty(\operatorname{Var}(\gamma_2) - \frac{I_{\gamma_1}}{\det(\mathbf{I}_{\boldsymbol{\gamma}})})}
	\end{equation}
	
\begin{figure}[H]
		\centering
		\includegraphics[width=\linewidth]{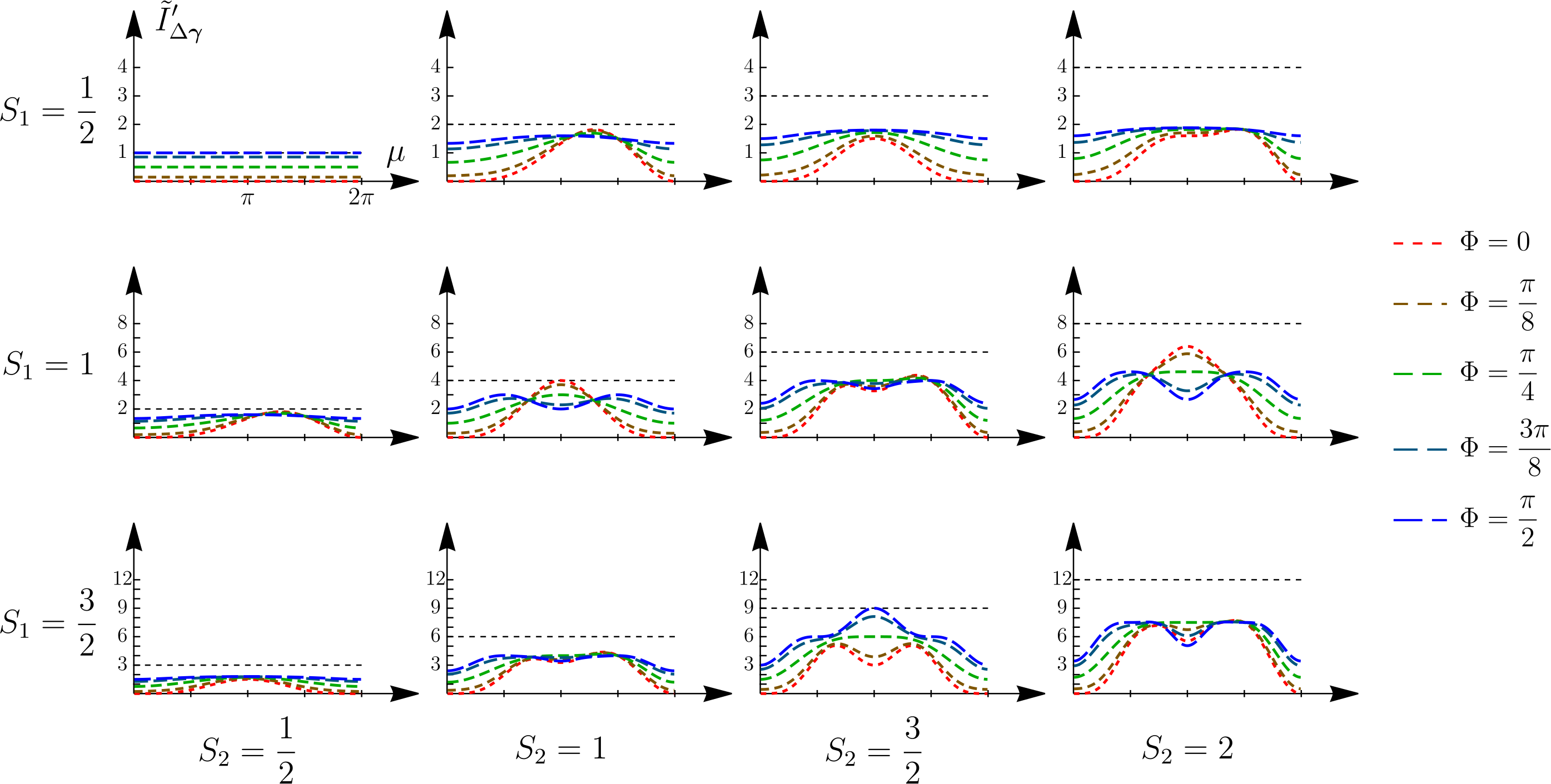}
		\caption{Plots of values of $\tilde{I}'_{\Delta \boldsymbol{\gamma}}$ defined in \eqref{eq:rotation_dif_2} for selected values of $S_1,\,S_2$ and $\Phi$. Note the dashed black line, corresponding to the value $\tilde{I}_{\Delta \boldsymbol{\gamma}} = 4S_1S_2$. In particular, note the asymmetry appearing when $S_2 - S_1$ is a half-integer.}
		\label{fig:qfirotationdifferencecorr}
	\end{figure}

	Based on this it is possible to consider another extreme case, where we allow one of the variances to saturate the bound, eg. $\operatorname{Var}(\gamma_1) = \frac{I_{\gamma_2}}{\det(\mathbf{I}_{\boldsymbol{\gamma}})}$, which implies $\operatorname{Cov}(\gamma_1, \gamma_2) = -\frac{(\mathbf{I}_{\boldsymbol{\gamma}})_{12}}{\det(\mathbf{I}_{\boldsymbol{\gamma}})}$, the bounds for variance of the rotation difference are changed to
	
	\begin{equation}\label{eq:rotation_dif_2}
	\operatorname{Var}(\gamma_2 - \gamma_1) = 
	\operatorname{Var}(\gamma_2) + \operatorname{Var}(\gamma_1) - 2 \operatorname{Cov}(\gamma_1,\gamma_2) \geq 
	\frac{I_{\gamma_1} + I_{\gamma_2} + 2 (\mathbf{I}_{\boldsymbol{\gamma}})_{12}}{\det(\mathbf{I}_{\boldsymbol{\gamma}})} \equiv 
	\tilde{I}_{\Delta \boldsymbol{\gamma}}^{'-1},
	\end{equation}
	which has been plotted with respect to $\mu$ in Fig. \ref{fig:qfirotationdifferencecorr} for several values of $S_1,\,S_2$ and $\Phi$.

	In this case we again see similarity between the plots from Fig. \ref{fig:QFI_Plots} and the equal-spin situations. This time, however, we note that there is an apparent asymmetry resulting in higher values of $\tilde{I}_{\Delta \boldsymbol{\gamma}}'$ for values of squeezing $\mu$ closer to $2\pi$ rather than $0$. This asymmetry would have been resolved by plotting with $\mu$ up to $4\pi$ and can be justified by a heuristic drawn from a single-spin squeezing. Note that squeezing for integer-spins is exactly periodic with period $\pi$, ie. state returns to itself when $\mu = \pi$. For half-integers the situation is different -- squeezing of $\mu = \pi$ yields a coherent state, but an antipodal one. Thus, a joint evolution of squeezed integer spin and a half-integer spin is essentially periodic with a period of $2\pi$. In this case we observe the same phenomenon, extended due to the use of two-spin instead of single-spin squeezing.

	\subsection{Axis angle estimation and  
          sum rule}
	
		As we have seen, the SRU protocol allows for a QFI for
                rotation angle estimation that scales quadratically
                with the number of systems for any equatorial axis of
                rotation. It is natural to ask whether it also
                provides an advantage in the task of rotation axis
                estimation. In this section we shortly note the
                properties of the SRU protocol found for this problem
                in the 2-spin setting. Closed expressions for the
                general case are hard to obtain, thus we will limit
                our considerations to a particularly compelling
                case. First, it is easy to demonstrate that the 
                scaling of the single-parameter QFI 
                for the estimation of the rotation axis angle $\Phi$
                depends on the angle of rotation $\gamma$, 
		
		\begin{align}
				I_\Phi\qty(\ket{\text{SRU}_2(\gamma,\Phi;\mu)}) & \,\propto \,\sin[2](\frac{\gamma}{2}),
		\end{align}
		suggesting that the case $\gamma = \pi$ is optimal. This is not the case as \mbox{$I_\Phi(\ket{\text{SRU}_2(\pi,\Phi;\mu)}) = 8S_M$}\\
		, scaling linearly with the size of the spin system. Consider then the intermediate case with the state turned by a quarter-turn, $\gamma = \pi/2$, which yields
		\begin{equation}\label{eq:axis_estimation_QFI}
		{\footnotesize
			\begin{aligned}
				I_\Phi(\ket{\text{SRU}_2(\sfrac{\pi}{2},\Phi;\mu)}) = S_M & \left(2 S_M+3 + \left(2 S_M-1\right) \cos (2 \Phi ) \cos ^{2 S_P}(\mu ) \right. \\
				& \left.-4 S_M \cos ^2(\Phi ) \cos ^{4 S_P}\left(\frac{\mu }{2}\right)\right)\,.
			\end{aligned}
		}
		\end{equation}
		This expression can be compared with a similar Eq.\eqref{eq:spin_QFI_anyAxis}. Their sum becomes fully independent from the choice of the rotation axis $\Phi$, 
		
		\begin{equation}
			I_\Phi(\ket{\text{SRU}_2(\sfrac{\pi}{2},\Phi;\mu)}) + I_\gamma(\ket{\text{SRU}_2(\gamma,\Phi;\mu)}) = 4 S_M \qty(S_M+1-S_M \cos ^{4 S_P}\qty(\frac{\mu }{2})),
		\end{equation}
		which shows a peculiar sum-rule: For a constant squeezing parameter, the sum of QFI for rotation angle and the axis angle depends only on the squeezing parameter $\mu$, which can be understood as a trade-off or an uncertainty relation. The better the sensitivity in the estimation of the rotation angle, the harder it gets to determine the axis of rotation. 
		
	\subsection{Compatibility of measurements in multivariate estimation problem}	
	
	\subsubsection{Compatibility of rotation angle and axis angle estimation}

		\rd{One may further ask for general attainability of the
                  Cramer-Rao bounds in a scenario where 
                  one intends to estimate $\gamma$ and $\Phi$
                  simultaneously, based on the commutation of SLDs 
                  on average 
                  as in eq.~\eqref{eq:SLD_comm_purestates}. Due to
                  the inaccessibility of a closed-formed expressions we
                  conducted a numerical investigation based for
                  selected values of the spin (equal for both spins,
                  $S_M = S_P = J$). 
                  The results                   are 
                  presented in Fig. \ref{fig:sldcommutativityplots}.
		\begin{figure}[H]
			\centering
			\includegraphics[width=.9\linewidth]{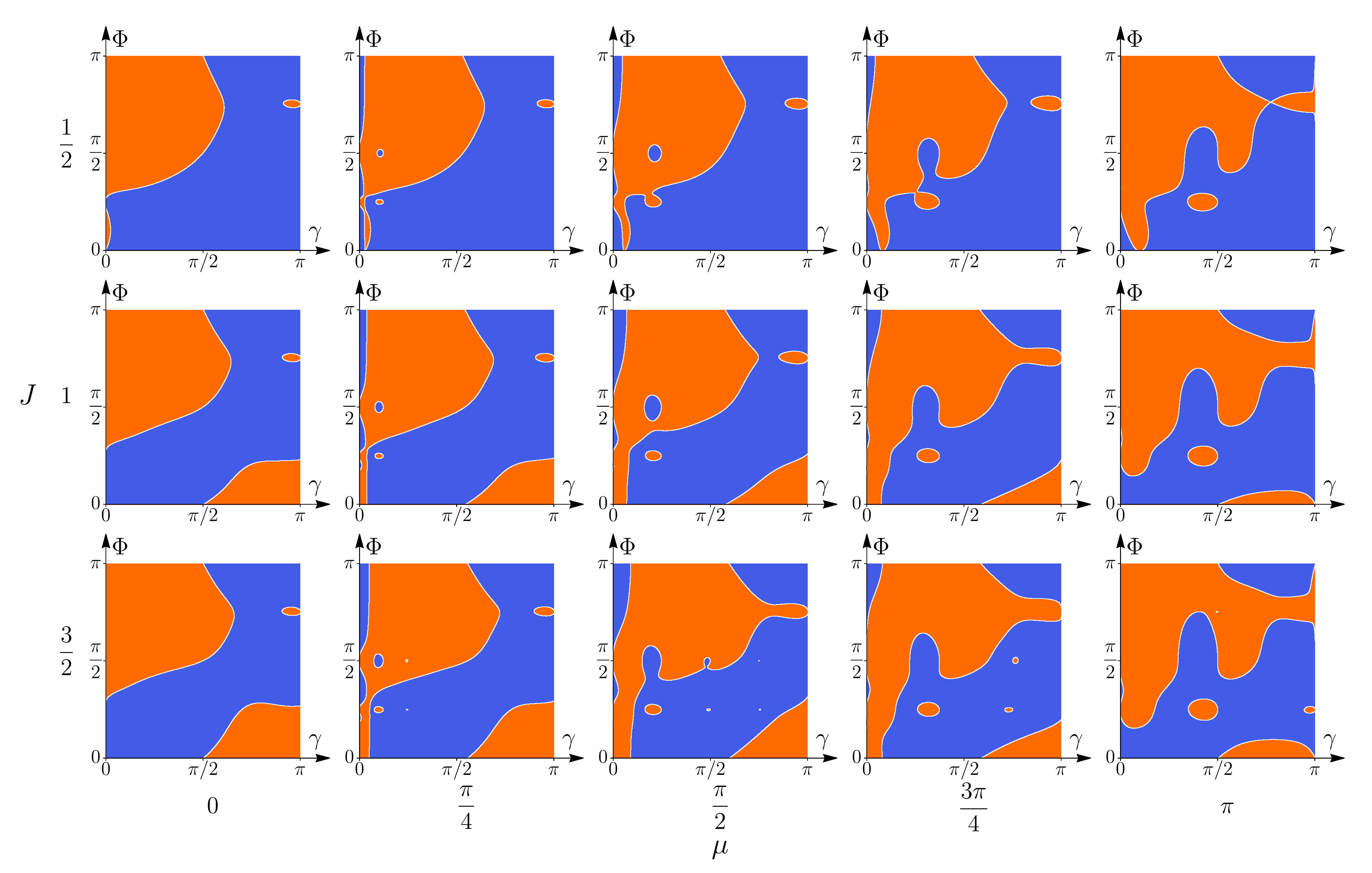}
			\caption{\rd{Plots depicting the signs of the expectation value of the commutator of SLDs $\Tr[\rho_\psi\comm{L_\gamma}{L_\Phi}]$ with the probe state $\ket{\psi} = \ket{\text{SRU}_2(\gamma,\Phi;\mu)}$ for a selection of values of the squeezing $\mu$ and the spin $J$ for both subsystems involved. Color coding is Red for positive and Blue for negative values. Note that for $J\geq1$ for every $\gamma = \text{const}$ one finds both positive and negative values, which entails at least one single zero
               (white border lines between the orange and blue regions)             and, in turn, saturability of the quantum Cramer-Rao bound.}}
			\label{fig:sldcommutativityplots}
		\end{figure}
		By looking at the plots we see one important feature
                -- there exist regions for which the SLDs commute 
                on average, 
                namely the 
                border lines between the orange and blue regions.
                This property is not exhibited by the SU(1,1)
                interferometry. 
                Furthermore, it is useful to note that for $J\geq 1$ for almost
                any value of $\gamma$ there exists a corresponding
                $\Phi$, for which the SLDs commute. Nevertheless, the
                structure of regions for which the SLDs commute 
                suggests 
                a need for numerical determination of the
                saturating measurement 
                in each case.  
	}

	\subsubsection{Compatibility of estimation 
of two noncommuting rotations}
	\rd{
		One may also consider the question of estimating
                rotations with respect to the $X$ and $Y$ axes 
                by angles $\gamma_x = \gamma \cos \Phi$ and $\gamma_y
                = \gamma \sin \Phi$, respectively.  This bears the
                closest similarity to the SU(1,1) displacement
                detection scheme, 
                with two rotations with orthogonal rotation directions
                in analogy to $p$ and $q$ quadratures. 
                As we are working with displacements from SU(2) group, we may implicitly define operators $B_x$ and $B_y$ as 
		\begin{equation}
			\begin{aligned}
				\ket{\partial_{\gamma_i}\psi} = \partial{\gamma_i}\ket{\text{SRU}_2(\sfrac{\pi}{2},\Phi;\pi)} = &
			i e^{-i \mu S_z\otimes S_z} B_i e^{i \gamma S(\Phi)\otimes\mathbb{I}} e^{i \mu S_z\otimes S_z}  \times \\
			& \times \ket{S_M, S_M}_{\pi/2,0}\otimes\ket{S_P, S_P}_{\pi/2,0}.
			\end{aligned}
		\end{equation}
		where $i = x,y$.
		It is easily shown that 
		\begin{equation}
			\Tr(\op{\psi}\comm{L_{\gamma_x}}{L_{\gamma_y}}) = 8\ev{\comm{B_x}{B_y}}{\psi}\,.
		\end{equation}
		Since each of the $B_i$ operators is proportional to a
                generator of the $SU(2)$ group, one can decompose the commutator into the sum of generators,
		\begin{equation}
			\comm{B_x}{B_y} = \qty(a_x S_x + a_y S_y + a_z S_z)\otimes\mathbbm{1}
		\end{equation}
		with
			\begin{align}
				a_x & = \frac{2 i \sin (\gamma ) \sin \qty(\frac{\gamma }{\sqrt{2}}) \sin (\phi ) \qty(\sqrt{2} \sin \qty(\frac{\gamma }{\sqrt{2}}) \cos ^2(\phi )+\gamma  \cos
					\qty(\frac{\gamma }{\sqrt{2}}) \sin ^2(\phi ))}{\gamma ^2} \\
				a_y & = \frac{2 i \sin (\gamma ) \sin ^2\qty(\frac{\gamma }{\sqrt{2}}) \qty(\sqrt{2}-\gamma  \cot \qty(\frac{\gamma }{\sqrt{2}})) \sin ^2(\phi ) \cos (\phi )}{\gamma ^2}\\
				a_z & = -\frac{2 i \sin ^2\qty(\frac{\gamma }{\sqrt{2}}) \qty(\sin (\gamma ) \cos ^2(\phi )+\cos (\gamma ) \cot \qty(\frac{\gamma }{\sqrt{2}}))}{\gamma }.
			\end{align}
		The general expression for the commutator is then given by
		\begin{equation}
			\ev{\comm{B_x}{B_y}}{\psi} = \frac{4S_M i \sin ^2\qty(\frac{\gamma }{2}) \cos \qty(\frac{\gamma }{2}) \cos ^{2S_M}\qty(\frac{\mu }{2}) \sin (\phi ) \qty[2 \sin \qty(\frac{\gamma }{2}) \cos ^2(\phi )+\gamma  \cos \qty(\frac{\gamma }{2}) \sin ^2(\phi )]}{\gamma ^2}
		\end{equation}
		Using simple trigonometric limits of the type $\lim_{x\rightarrow0} \frac{\sin(x)}{x} = 1$ one finds that 
		\begin{equation}
			\lim_{\gamma \rightarrow 0} \comm{B_x}{B_y} = -i\sqrt{2}S_z.
		\end{equation}
		Thanks to this and the alignment of the initial state
                with the $X$ axis we can say that for 
                rotations
                with rotation angles close to 
                $\gamma = 0$ the SLDs commute on average,
		\begin{equation}
			\lim_{\gamma \rightarrow 0}\Tr(\op{\psi}\comm{L_{\gamma_x}}{L_{\gamma_y}}) =
			8\lim_{\gamma \rightarrow 0} \ev{\comm{B_x}{B_y}}{\psi} = 0\,,
		\end{equation}
		which guarantees the existence of a measurement that saturates the multivariate QCRB.
 	}
		
	\subsection{Axis estimation in the context of estimation with time reversal}
		Consider Eq.\eqref{eq:axis_estimation_QFI} and observe that by setting $\mu = \pi$ we get the maximal value of the sum, which can be also re-expressed in terms of the number $N_M = 2S_M$ of half-spin subsystems,
		\begin{equation}
			\max_\Phi I_\Phi(\ket{\text{SRU}_2(\sfrac{\pi}{2},\Phi;\pi)}) = 4S_M^2 + 2 S_M = N_M^2 + N_M
		\end{equation}
		which would suggest exceeding the Heisenberg limit of $N_M^2$. This is not the case, as one can rewrite the rotation generated by $S(\Phi)$ as
		\begin{equation}
			e^{i \gamma S(\Phi)} = e^{-i \Phi S_z} e^{i \gamma S_y} e^{i \Phi S_z},
		\end{equation}
		to notice that the rotation of the angle $\Phi$ is
                imprinted twice on the state. One thus deviates from
                the standard protocol that leads to the HL. The double
                imprinting comes with a slight twist: 
                the evolution 
                is first forward in time and the second time 
                backward in time. In other words, we have 
		double imprinting of $\Phi$ with time reversal. 
		This can be also interpreted as estimation of the parameter of a Hamiltonian evolution acting as a superoperator on a different, fixed Hamiltonian evolution.
	
		We wish to calculate the optimal value of Fisher information $I_\gamma(\ket{\psi_\gamma})$ from the state $\ket{\psi_\gamma}$ and its derivative given by
	
		\begin{align}
			\ket{\psi_\gamma} = e^{i\gamma H_0} e^{iH_1} e^{-i\gamma H_0}\ket{\psi_0} && \Longrightarrow&&
			\partial_\gamma \ket{\psi_\gamma} = i \comm{H_0}{e^{i\gamma H_0} e^{iH_1} e^{-i\gamma H_0}}\ket{\psi_0}.
		\end{align}
	Assume that $H_0$ is a constant Hamiltonian rescaled in such a way that its eigenvalues satisfy $\lambda_{max} = -\lambda_{min} = N$, and the free parameters are the input state $\ket{\psi_0}$ and the auxiliary Hamiltonian $H_1$. Inspection of these expressions shows that $I_\gamma(\ket{\psi_\gamma}) = 0$ whenever $\comm{H_0}{e^{i H_1}}= 0$. This suggests that the largest value of $I$ should be connected to $\text{argmax}_{H_1}\qty(\abs{\comm{H_0}{e^{i H_1}}})$ in a suitable metric. 
		
	In order to bound the optimal value of $I_\gamma(\ket{\psi_\gamma})$ from below we introduce an Ansatz of $\ket{\psi_0} \propto \ket{\lambda_{max}} + e^{i \alpha} \ket{\lambda_{min}}$ and assume that the auxiliary Hamiltonian $H_1$ does not mix the subspace $\text{span}\qty(\ket{\lambda_{min}},\ket{\lambda_{max}})$ with its complement. Explicit calculations in Appendix \ref{app:superchannel_estim} show that by setting $e^{i H_1} = \cos \chi \sigma_x + \sin \chi \sigma_y$ on the two-dimensional subspace mentioned above we find the maximum of the Fisher information,
	\begin{equation}
		\max_{H_1, \ket{\psi_0}} I_\gamma(\ket{\psi_\gamma}) = 4N^2\,,
	\end{equation}
	which is well above the value of $N^2 + N$ demonstrated for the axis estimation problem. 
		
%
		
	\subsection{Numerical study of two-axis squeezing}
	
		For completeness, we shortly consider a variant of the
                SRU protocol where instead of one-axis squeezing we
                utilize the two-axis squeezing operator. The full
                protocol 
                is now defined as
		\begin{equation}
			\ket{\text{SRU}^2_2(\gamma,\Phi;\mu)} = e^{-\mu(S_-^2 - S_+^2)}e^{i \gamma S(\Phi)}e^{\mu(S_-^2 - S_+^2)}\ket{S,S}  \label{eq:2_axis_SRU_state},
		\end{equation}
		in analogy to the SRU protocol 
                defined in \eqref{eq:1_spin_SRU}. Here the squeezing Hamiltonian has been changed to $i (S_+^2 - S_-^2)$ and the input state is a spin coherent state pointing along the $Z$ axis, following the preparation presented in \cite{KU93}. Therein, the authors have demonstrated that the variance of a spin in a selected direction orthogonal to the main axis of the squeezed state for large spins approaches a constant. Later the two-axis squeezing has been studied theoretically \cite{KW15} and only recently an experimental realization has been proposed \cite{HEtAl22}.  
		
		In this case analytical treatment is not accessible.
                Therefore we present a set of plots for selected
                values of $S$ which present the qualitative behaviour
                of two-axis squeezing. Despite clearly periodic
                behaviour for $S = 1$ and $S = \sfrac{5}{2}$, for
                larger spins the QFI fails to reach the value of $4S^2$ demonstrated to be attainable for one-axis squeezing. Furthermore, despite isolated points at which Fisher information agrees for different rotation axes, no stable plateaus can be found.
		\begin{figure}[H]
			\centering
			\includegraphics[width=.75\linewidth]{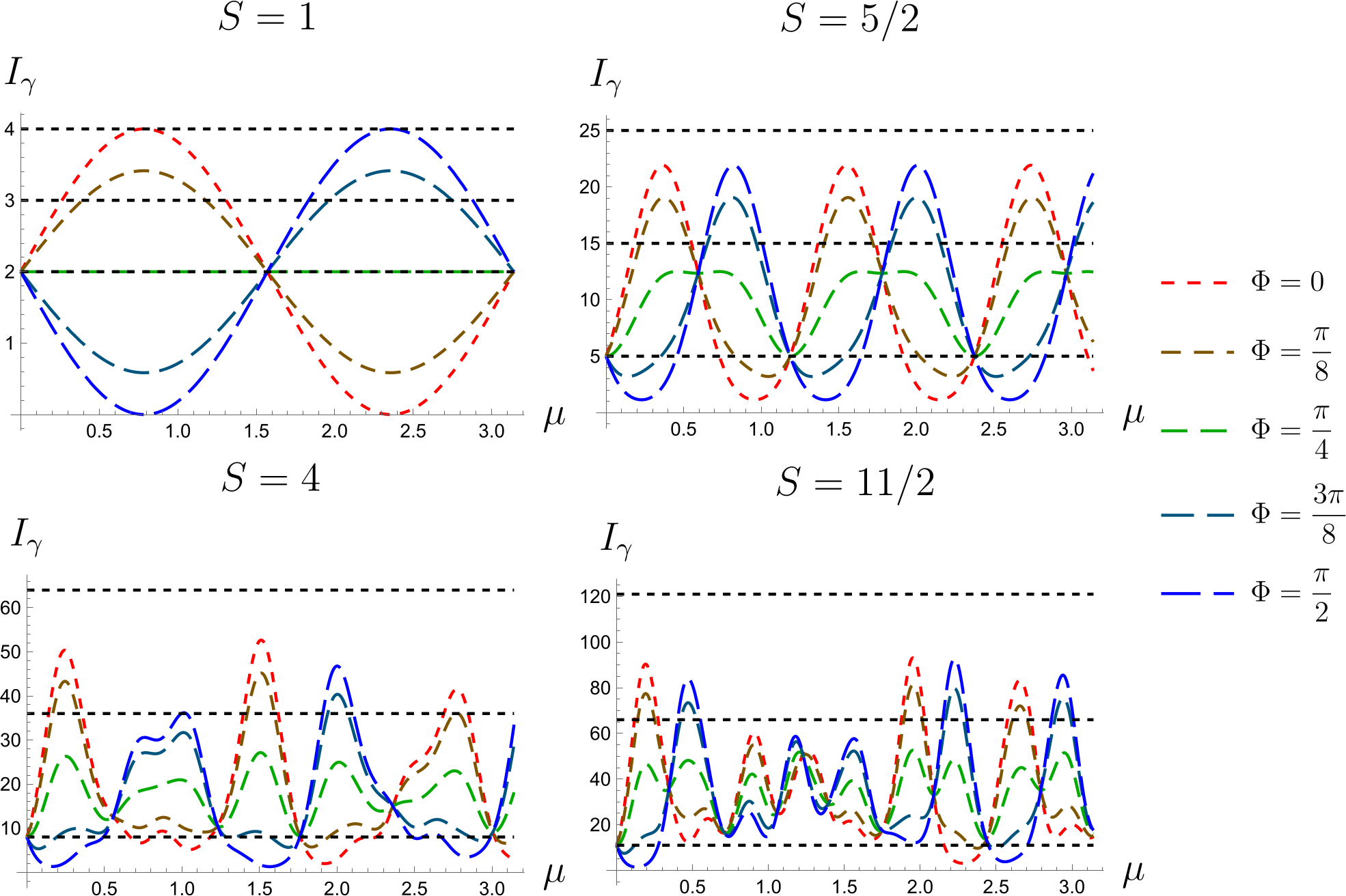} 
			\caption{
				Plots of quantum Fisher information $I_\gamma(\ket{\text{SRU2}(\gamma,\Phi;\mu)}_1)$ for state $\ket{\text{SRU2}(\gamma,\Phi;\mu)}_1$ prepared with the two-axes squeezing protocol \eqref{eq:2_axis_SRU_state}. Note the aperiodic behaviour and absence of stable plateaus.}
			\label{fig:QFI_Plots_1spin_2axes}
		\end{figure}

	\section{Concluding Remarks}\label{sec:conclude}

	In this work we have translated $\text{SU}(1,1)$ interferometry,
        understood 
        by Caves as a displacement detector \cite{Caves20}, to the field
        of spin-rotation estimation. The introduced
        squeeze-rotate-unsqueeze protocol for two- and one-spin
        one-axis squeezing, analogous to the
        squeeze-displace-unsqueeze procedure applied in the $\text{SU}(1,1)$
        interferometry, has been demonstrated to provide an advantage
        over the %
        best classical scheme. %
        This is evidenced by the Quantum Fisher Information (QFI)
        approaching a scaling with the square of the size of the
        system (``the Heisenberg scaling''), as opposed to the linear
        scaling of the Standard Quantum Limit. 

        Most importantly, for the specific intermediate value of squeezing, $\mu = \pi/2$, we have found 
        simultaneous quadratic scaling {of the single-parameter values
          of the QFI} for the estimation of the rotation %
        angles %
        around 
        {\it all} equatorial axes, which can be compared to the
        simultaneous
        {amplification of displacements}
        in an arbitrary direction in the phase space for
        bosonic systems. 
        The effect is robust in the sense of the existence of a plateau around the specific squeezing parameter, which grows with the size of the probe system. 
	Furthermore, we have shown that the introduced scheme provides the advantage of quadratic scaling for rotation-axis estimation given control over the rotation angle. 
        Finally, to provide a more complete picture we have demonstrated the behaviour of the QFI for the SRU protocol with two-axis squeezing, showing lack of crucial properties like saturation of Heisenberg limit or appearance of plateaus.\\
	
	The SRU protocol for a single-spin can be seen as a special case of the delta-kick squeezing (DKS) protocol introduced in \cite{CGSP21}. One has, however, to stress that in our work the inspiration comes from the SU(1,1) displacement detector's ability to achieve an advantage for arbitrary direction of displacement. Therefore, we have focused on finding stable advantages over a wide range of different axes of rotation. The 2-spin SRU protocol goes beyond DKS, 
        in the sense that the size of the probe spin 
        can be used to control the specific 
        dependence of the QFI on the squeezing parameter $\mu$, allowing an extension of the plateau of quadratic scaling for an arbitrary rotation axis in the equatorial plane.
        This allows the 
        estimation of different
        components of a magnetic field that lead to non-commuting
        rotations 
        with Heisenberg-limited sensitivity.
        Future research could extend our investigations 
        in the spirit of \cite{KEtAl19} to ask for variational generation of spin states which are flexible for rotation-angle estimation over a wide range of axes.
Several other open directions are left to pursue in the context of the 2-spin SRU protocol.
         For instance, one may consider the QFI of the probe system
         for estimating the rotation angle of the primary system. Another future direction involves seeing joint measurability of noncommuting rotations in the perspective of recent attempts to incorporate Heisenberg uncertainty relations into optimization of measurement errors \cite{LW21}.
         
         Finally, it is interesting to mention a recent experimental effort to generate a four photon state with $J=3/2$ and tetrahedral symmetry in the polarization-related $SU(2)$ representation \cite{ferretti2023generating}. Although not scalable towards larger $J$, it is an example for efforts towards effective schemes for sensing rotations, even beyond actual spin systems. In this sense our method may prove useful for equivalents of rotations in polarization of multi-photon states using methods for polarization-squeezing \cite{chirkin_1993, klyshko97polarization, korolkova02, schnabel03}.
	
	\section*{Acknowledgments}
		JCz and KŻ would like to thank Jan Kołodyński and Rafał Demkowicz-Dobrzański for fruitful discussion, in particular on the topic of multiparameter estimation. 
		This work realized within the DQUANT  QuantERA II Programme 
		was supported by the National Science Centre, Poland, under the contract number 2021/03/Y/ST2/00193.
		KŻ and JCz acknowledge gratefully financial support by Narodowe Centrum Nauki under the Preludium BIS grant number 2019/35/O/ST2/01049.
	
	\appendix
	\section{Quantum Fisher information for SDU protocol for one- and two-mode scenario}\label{app:SDU_QFI}
	
	Based on the expressions provided in \cite{Caves20} it is
        rudimentary to show that the state of a single-mode bosonic
        system undergoing the SDU protocol is given by
	\begin{equation}
		\ket{\psi_1(\gamma,r)} = S_1^\dagger(r)D(\gamma)S_1(r)\ket{0} = \ket{\alpha e^r + i \beta e^{-r}}
	\end{equation}
	where $\gamma = \alpha = i\beta$ and we consider $\ket{x}$ to be a coherent state with displacement equal to $x$. QFI with respect to $\alpha$ and $\beta$ is then given by
	\begin{align}
		\vb{I}(\ket{\psi_1(\gamma,r)}) = 4\mqty(e^{2r}&0\\0&e^{-2r}), &&
		\Tr(\rho_{\psi_1}\comm{L_\alpha}{L_\beta}) = 8i\alpha\beta
	\end{align}
	where $\vb{I}_{11}(\cdot) = I_\alpha(\cdot)$ and $\vb{I}_{22}(\cdot) = I_\beta(\cdot)$. We see that the QFIs are inversely proportional and the QCRB is saturable only when either $\alpha = 0$ or $\beta = 0$, ie. one of the displacements is estimated around 0.
	
	The two-mode input state is given by
	\begin{equation}\label{eq:2_beam_preparation}
	\ket{\psi_2(\gamma,r)} = S_2^\dagger(r)\qty(D(\gamma)\otimes\mathbbm{1})S_2(r)\qty(\ket{0}\otimes\ket{0}) = B \qty(\ket{\alpha e^r + i \beta e^{-r}}\otimes\ket{\alpha e^{-r} + i \beta e^r})
	\end{equation}
	where $B = e^{i J_2 \pi/2}$ is the operator representation of a beamsplitter.
	In this case considerations of QFI and of average SLD commutation yield
	\begin{align}
		\vb{I}(\ket{\psi_2(\gamma,r)}) = 8\mqty(\cosh(2r)&0\\0&\cosh(2r)), &&
		\Tr(\rho_{\psi_1}\comm{L_\alpha}{L_\beta}) = 16i
	\end{align}
	and hence, we find that (a) any direction of the form $\cos(x)\alpha +\sin(x)\beta$ has the same QFI of the form $8\cosh(2r)$ and (b) one cannot saturate the bound above for two directions at once using a single measurement.

	The above statements may be seen as contradicting the prior results presented in \cite{Caves20}, where it was shown that it is possible to estimate both quadratures better than SNL \rd{check} sensitivity. However, it is not simultaneous in the sense of saturating the matrix-valued QCRB, but in the sense of saturating two single-parameter QCRBs independently. Using the following set of identities
	
	\begin{align*}
		B \qty(S_1\otimes S_1^\dagger)B^\dagger = S_2, &&
		B D(\gamma)^{\otimes 2} B^\dagger = \mathbbm{1}\otimes D(\sqrt{2}\gamma))
	\end{align*}
	\begin{align*}
		B(x_a+ip_b)B^\dagger = \qty(\qty(x_a+x_b) + i\qty(p_b - p_a))/\sqrt{2}
	\end{align*}
	Caves shows that the measurement of Bell variables $x_a + x_b$ and $p_a - p_b$ is actually equivalent to introducing a beamsplitter operator $B$, measuring single quadratures independently $x_a$ on the first mode and $p_b$ on the second, and applying beamsplitter $B^\dagger$ on the post-measurement state. With the preparation \eqref{eq:2_beam_preparation} it is clear that one implements an effective equivalent of two local measurements on two states undergoing independent SDU protocols.
	
	\section{Quantum Fisher information for SRU protocol with selected rotation axes and maximal squeezing} \label{app:1spin_selected_ax}
	
		We start with the expression for the state after the single-spin SRU protocol with maximal squeezing $\mu = \sfrac{\pi}{2}$,
		\begin{equation}
			\ket{\text{SRU}_1(\gamma,\Phi;\sfrac{\pi}{2})} =  e^{-i \frac{\pi}{2} S_z^2} e^{i \gamma S(\Phi)} e^{i \frac{\pi}{2} S_z^2} \ket{S, S}_{\pi/2,0}
		\end{equation}
		with $S(\Phi) = \cos\Phi S_x + \sin\Phi S_y$. We limit ourselves to $\Phi\in\qty{0,\frac{\pi}{2}}$. In order to arrive at closed expressions we split the presentation into the cases of integer and half-integer spin. Beginning with the integer case, we bring up a useful formula
		\begin{equation*}
			e^{i \frac{\pi}{2} m^2} = \frac{1+i}{2} + (-1)^m \frac{1-i}{2} = \frac{1+i}{2}\qty(1 -i(-1)^m),
		\end{equation*} 
		which shows that for integer spins
		\begin{equation}
			e^{i\frac{\pi}{2}S_z^2} = \frac{1+i}{2}\qty(\mathbb{I} - i e^{i \pi S_z}).
		\end{equation}
	
		Making use of this formula one can rewrite the squeezed states,
		
		\begin{align}
			e^{i\frac{\pi}{2}S_z^2} \ket{S,S}_{\pi/2,0} & = \frac{1}{2^S}\sum_{M = -S}^{S} \sqrt{\binom{2S}{S+M}} e^{i \frac{\pi}{2} M^2} \ket{S,M} \\
			& = \frac{1+i}{2}\frac{1}{2^S}\sum_{M = -S}^{S} \sqrt{\binom{2S}{S+M}} \qty(1 - i(-1)^M) \ket{S,M} \\
			& = \frac{1+i}{2} \qty(\ket{S,S}_{\pi/2,0} - (-1)^S i \ket{S,S}_{\pi/2,\pi}),
		\end{align}
		which is clearly a NOON state. Now we consider the case $\Phi = 0$, for which we have $S(0) = S_x$, such that the rotation $e^{i\gamma S_x}$ introduces phases to the components,
		
		\begin{align}
			e^{i \gamma S_x} e^{i\frac{\pi}{2}S_z^2} \ket{S,S}_{\pi/2,0} & = \frac{1+i}{2} \qty(e^{i S \gamma}\ket{S,S}_{\pi/2,0} - (-1)^S i e^{-i S \gamma}\ket{S,S}_{\pi/2,\pi}).
		\end{align}
		In the last step we need an additional intermediate calculation,
			\begin{align}
				e^{-i\frac{\pi}{2}S_z^2} \ket{S,S}_{\pi/2,\pi} & = \frac{1}{2^S}\sum_{M = -S}^{S} \sqrt{\binom{2S}{S+M}} e^{i \frac{\pi}{2} M^2} (-1)^{S-M} \ket{S,M} \\
				& = \frac{1-i}{2}\frac{1}{2^S}\sum_{M = -S}^{S} \sqrt{\binom{2S}{S+M}} \qty((-1)^{S-M} + i(-1)^S) \ket{S,M} \\
				& = \frac{1-i}{2} \qty(\ket{S,S}_{\pi/2,\pi} + (-1)^S i \ket{S,S}_{\pi/2,0}),
			\end{align}
		from which we finally write
		\begin{align}
			\ket{\text{SRU}_1(\gamma,0;\sfrac{\pi}{2})} & = \frac{1}{2}\qty[\qty(e^{iS\gamma} + e^{-iS\gamma}) \ket{S,S}_{\pi/2,0} + (-1)^Si\qty(e^{iS\gamma} - e^{-iS\gamma})\ket{S,S}_{\pi/2,\pi}] \\
			& = \cos(S\gamma) \ket{S,S}_{\pi/2,0} - (-1)^S\sin(S\gamma)\ket{S,S}_{\pi/2,\pi}.
		\end{align}
	
		The case of $S(\sfrac{\pi}{2}) = S_y$ is easily resolved by using two formulas,
		\begin{align}
			e^{i\gamma S_y} \ket{S,S}_{\theta,0} = \ket{S,S}_{\theta-\gamma,0}, && e^{i\frac{\pi}{2}S_z^2} \ket{S,S}_{\theta,\phi} = \frac{1+i}{2}\qty(\ket{S,S}_{\theta,\phi} - (-1)^S i \ket{S,S}_{\theta,\phi+\pi}),
		\end{align}
		which lead directly to the expression,
		\begin{equation}
			{\footnotesize\ket{\text{SRU}_1(\gamma,\sfrac{\pi}{2};\sfrac{\pi}{2})}  = \ket{S,S}_{\pi/2-\gamma,0} + \ket{S,S}_{3\pi/2-\gamma,\pi} + (-1)^S i \qty(\ket{S,S}_{\pi/2-\gamma,0} - \ket{S,S}_{3\pi/2-\gamma,0})}.
		\end{equation}
	
		Expressions for half-integer spin are obtained in a similar way with the formula,
		\begin{equation}
			e^{i\frac{\pi}{2}\qty(\frac{2n+1}{2})^2} = \sqrt{2}e^{i\frac{\pi}{8}}\cos(\frac{\pi}{2}\frac{2n+1}{2}),
		\end{equation}
		which yields a similar form of the maximal squeezing operator,
		\begin{equation}
			e^{i\frac{\pi}{2}S_z^2} = \frac{e^{i\frac{\pi}{8}}}{\sqrt{2}}\qty(e^{i\frac{\pi}{2}S_z} + e^{-i\frac{\pi}{2}S_z}).
		\end{equation}

	\section{QFI for rotation angle of 2-spin SRU around any equatorial axis}\label{app:2spin_equat_axes}
	
		The starting point is the 2-spin state after the SRU protocol,
		
		\begin{equation}
			\ket{\text{SRU}_2(\gamma,\Phi;\mu)} = e^{-i \mu S_z\otimes S_z} e^{i \gamma S(\Phi)\otimes\mathbb{I}} e^{i \mu S_z\otimes S_z} \qty(\ket{S_M, S_M}_{\pi/2,0}\otimes\ket{S_P, S_P}_{\pi/2,0}),
		\end{equation}
		which in this case turns out to be a simple expression to handle. For brevity we shall write  $\ket{j,k}\equiv\ket{S_M,j}\otimes\ket{S_P,k}$. Making use of this notation consider the squeezed state,
		
		{\footnotesize
			\begin{equation}
				\begin{alignedat}{3}
					&\ket{\text{SRU}_2(\gamma,\Phi;\mu)}& = \frac{1}{2^{S_M + S_P}}\sum_{j=-S_M}^{S_M} \sum_{k = -S_P}^{S_P}
					\sqrt{\binom{2S_M}{S_M - j}\binom{2S_P}{S_P - k}}
					&\qty(\cos\frac{\gamma}{2} + e^{i(\Phi-k \mu)}\sin\frac{\gamma}{2})^{S_M + j} \nonumber& \\
					&&& \qty(\cos\frac{\gamma}{2} - e^{-i(\Phi-k \mu)}\sin\frac{\gamma}{2})^{S_M - j} \ket{j,\,k},&
				\end{alignedat}
		\end{equation}}
		and its derivative,
			{\small\begin{equation}
				\begin{aligned}
					\ket{\partial_\gamma\text{SRU}_2(\gamma,\Phi;\mu)} = & \frac{1}{2^{S_M + S_P}}
					 \sum_{j=-S_M}^{S_M} \sum_{k = -S_P}^{S_P}\sqrt{\binom{2S_M}{S_M - j}\binom{2S_P}{S_P - k}}\\
					& \times \qty(\frac{(S_M-j) \cos (k \mu -\Phi )-S_M (\sin (\gamma )+i \cos (\gamma ) \sin (k \mu -\Phi ))}{\cos (\gamma )-i \sin (\gamma ) \sin (k \mu -\Phi )}) \\
					& \times \qty(\cos\frac{\gamma}{2} + e^{i(\Phi-k \mu)}\sin\frac{\gamma}{2})^{S_M + j}\qty(\cos\frac{\gamma}{2} - e^{-i(\Phi-k \mu)}\sin\frac{\gamma}{2})^{S_M - j} \ket{j,\,k}
				\end{aligned}
			\end{equation}}
		In order to simplify matters it is useful to substitute $j$ by $S_M - j$ and consider auxiliary sums of the form,
			\begin{equation} \label{eq:F_expressions}
				{\small\begin{alignedat}{4}
					&F_0 =&& \frac{1}{2^{2S_M}}\sum_{j = 0}^{2S_M}&     \binom{2S_M}{S_M -j} (1+a)^{2S_M - j} (1-a)^j & = 1, \\
					&F_1 =&& \frac{1}{2^{2S_M}}\sum_{j = 0}^{2S_M}& j   \binom{2S_M}{S_M -j} (1+a)^{2S_M - j} (1-a)^j & = S_M (1-a), \\
					&F_2 =&& \frac{1}{2^{2S_M}}\sum_{j = 0}^{2S_M}& j^2 \binom{2S_M}{S_M -j} (1+a)^{2S_M - j} (1-a)^j & = S_M\qty(S_M (1-a)^2 + \frac{1 - a^2}{2}),
				\end{alignedat}}
			\end{equation}
		with $a = \cos(\Phi - k\mu)\sin(\gamma)$. With these one can express the relevant inner products as
		
			\begin{equation}
				{\scriptsize\begin{aligned}
					\braket{\text{SRU}_2(\gamma, \Phi;\mu)}{\partial_\gamma\text{SRU}_2(\gamma,\Phi;\mu)} & = \!\!\!\!\sum_{k=-S_P}^{S_P}\!\!\!\! \frac{\binom{2S_P}{S_P - k}}{2^{2S_P}}\qty[\qty(F_0\; F_1)
					\cdot\mqty(
						\frac{S_M (i \text{c} (2 \gamma) \text{s} (\Phi-k \mu)+\text{c} (\Phi-k \mu)-\text{s} (2 \gamma))}{i \text{s} (2 \gamma) \text{s} (\Phi-k \mu)-\text{s} ^2(\gamma)+\text{c} ^2(\gamma)} \\
						-\frac{\text{c} (\Phi-k \mu)}{i \text{s} (2 \gamma) \text{s} (\Phi-k \mu)-\text{s} ^2(\gamma)+\text{c} ^2(\gamma)}
					)], \\
					\braket{\partial_\gamma\text{SRU}_2(\gamma,\Phi;\mu)} & = \!\!\!\!\sum_{k=-S_P}^{S_P}\!\!\!\! \frac{\binom{2S_P}{S_P - k}}{2^{2S_P}}\qty[\qty(F_0\; F_1\; F_2) 
					\cdot \mqty(
						S_M^2 \left(\frac{2}{\text{s} (2 \gamma) \text{c} (\Phi-k \mu)+1}-1\right) \\
						\frac{2 S_M \text{c} (\Phi-k \mu) (\text{s} (2 \gamma)-\text{c} (\Phi-k \mu))}{\text{s} ^2(2 \gamma) \text{s} ^2(\Phi-k \mu)+\text{c} ^2(2 \gamma)} \\
						\frac{\text{c} ^2(\Phi-k \mu)}{\text{s} ^2(2 \gamma) \text{s} ^2(\Phi-k \mu)+\text{c} ^2(2 \gamma)}
					)],
				\end{aligned}}
			\end{equation}
		where for brevity we replaced $\text{c} \equiv \cos$ and $\text{s} \equiv \sin$.
		Standard algebraic transformations lead to,
		\begin{equation} \label{eq:F_products}
			{\small\begin{aligned}
				\braket{\text{SRU}_2(\gamma, \Phi;\mu)}{\partial_\gamma\text{SRU}_2(\gamma,\Phi;\mu)} & = i S_M \cos[2S_P](\frac{\mu}{2}) \sin\Phi, \\
				\braket{\partial_\gamma\text{SRU}_2(\gamma,\Phi;\mu)} & =\frac{S_M}{4} \qty(1 + 2S_M - (2S_M -1)\cos(2\Phi) \cos[2S_P](\mu)),
			\end{aligned}}
		\end{equation}
		which yields the expression \eqref{eq:spin_QFI_anyAxis}.
		
	\section{QFI for rotation angle of 1-spin SRU around any equatorial axis}\label{app:1spin_equat_axes}
	
		In order to calculate the Fisher information for a rotation with respect to an arbitrary  equatorial axis for a single spin we resort to the approach of first differentiating the operator and then evaluating its expectation value -- an approach that yields less knowledge about the states used to perform the task. Consider that
		
		\begin{equation}
			\partial_\gamma e^{-i\mu S_z^2} e^{i \gamma S(\Phi)} e^{i\mu S_z^2} = i e^{-i\mu S_z^2} e^{i \gamma S(\Phi)} S(\Phi) e^{i\mu S_z^2},
		\end{equation}
		which translates to
		\begin{equation}
			\begin{aligned}
				\braket{\text{SRU}_1(\gamma, \Phi;\mu)}{\partial_\gamma\text{SRU}_1(\gamma,\Phi;\mu)} & = i\ev{S(\Phi)}{S;\mu}, \\
				\braket{\partial_\gamma\text{SRU}_1(\gamma,\Phi;\mu)} & = \ev{S^2(\Phi)}{S;\mu},
			\end{aligned}
		\end{equation}
		and we recall from \eqref{eq:squeezed_spin_sx} that $\ket{S;\mu} = e^{-i\mu S_z^2}\ket{S,S}_{\pi/2,0}$. In this case it is helpful to write out the operators as $S(\Phi) = \frac{1}{2}\qty(e^{-i\Phi}S_+ + e^{i\Phi}S_-)$. Then one obtains
		\begin{align}
			\ev{S(\Phi)}{S;\mu} = & \frac{1}{2^{2S+1}}\sum_{M,M' = -S}^S \sqrt{\binom{2S}{S+M}\binom{2S}{S+M'}} e^{i \mu (M^2 - M'^2)} \nonumber \\
			& \times \mel{SM'}{e^{-i\Phi}S_+ + e^{i\Phi}S_-}{SM}\\
			 = & \frac{S}{2^{2S}} \sum_{M = -S}^S \binom{2S-1}{S+M}e^{-i \mu (2M+1) - i\Phi} + \binom{2S-1}{S+M-1}e^{i \mu (2M-1) + i\Phi} \\
			 = & \frac{S}{2^{2S}}\qty(e^{-i\Phi} \sum_{M=0}^{2S-1} \binom{2S-1}{M} e^{-i\mu M}e^{i\mu(2S-1-M)} + \text{c.c}) \\
			 = & S \cos\Phi \cos^{2S-1} \mu.
		\end{align}
	
	The second inner product can be calculated in a similar manner,
	\begin{equation}
		{\small\ev{S^2(\Phi)}{S;\mu} = \frac{1}{4}\qty(\ev{\qty(e^{-2i\Phi} S_+^2 + e^{2i\Phi} S_-^2)}{S;\mu} + \ev{\qty(S_+S_- + S_-S_+)}{S;\mu})},
	\end{equation}
	with the first term evaluating to
	
	\begin{equation}
		{\small\begin{aligned}
			\frac{1}{4}\ev{\qty(e^{-2i\Phi} S_+^2 + \text{h.c.})}{S;\mu} & = \frac{S(2S-1)}{2^{2S+1}} \qty(e^{2i\Phi}\sum_{M = 0}^{2S - 2} \binom{2S-2}{M}e^{2i\mu \qty(M - (2S-2-M))} + \text{c.c.}) \\
			& = \frac{S(2S-1)}{4} \cos(2\Phi)\cos[2S-2](2\mu).
		\end{aligned}}
	\end{equation}
	The second one
	\begin{equation}
		\frac{1}{4} \ev{\qty(S_+S_- + S_-S_+)}{S;\mu} = \frac{1}{2^{2S+1}}\sum_{M=0}^{2S}\binom{2S}{M}\qty((2M+1)S - M^2) = \frac{S(2S+1)}{4},
	\end{equation}
	which, collected together, yield the expression \eqref{eq:QFI_1spin} for $I_\gamma(\ket{\text{SRU}_1(\gamma, \Phi;\mu)})$.
	
	\section{Derivation of QFI for axis estimation}
	
	The starting point of this derivation is the calculation of the derivative of the state $\ket{\text{SRU}_2(\gamma,\Phi;\mu)}$ with respect to the axis angle $\Phi$,
	
	\begin{equation}
		{\footnotesize\begin{alignedat}{3}
			&\ket{\partial_\Phi\text{SRU}_2(\gamma,\Phi;\mu)}= && \frac{i \sin\frac{\gamma}{2}}{2^{S_M + S_P}}\sum_{j=-S_M}^{S_M} \sum_{k = -S_P}^{S_P}
			\sqrt{\binom{2S_M}{S_M - j}\binom{2S_P}{S_P - k}}
			  & \\
			&&& \times\qty(
			\frac{S_M + j}{e^{-i(\Phi -  k \mu )}\cos \left(\frac{\gamma }{2}\right) + \sin \left(\frac{\gamma }{2}\right)} + 
			\frac{S_M - j}{e^{i (\Phi-k\mu) } \cos \left(\frac{\gamma }{2}\right)-\sin \left(\frac{\gamma }{2}\right)}) &\\
			&&&\times\qty(\cos\frac{\gamma}{2} + e^{i(\Phi-k \mu)}\sin\frac{\gamma}{2})^{S_M + j} \qty(\cos\frac{\gamma}{2} - e^{-i(\Phi-k \mu)}\sin\frac{\gamma}{2})^{S_M - j}  \ket{j,\,k}.&
		\end{alignedat}}
	\end{equation}
	The prefactor $\sin\gamma/2$ suggests that the maximum of QFI may occur at $\gamma = \pi$. However, by explicit calculation we find that the derivative reads
	
	\begin{equation}
		\begin{alignedat}{3}
			&\ket{\partial_\Phi\text{SRU}_2(\pi,\Phi;\mu)}= && \frac{2 i}{2^{S_M + S_P}}\sum_{j=-S_M}^{S_M} \sum_{k = -S_P}^{S_P}
			\sqrt{\binom{2S_M}{S_M - j}\binom{2S_P}{S_P - k}} j
			& \\
			&&&\times\qty(e^{i(\Phi-k \mu)})^{S_M + j} \qty(- e^{-i(\Phi-k \mu)})^{S_M - j}  \ket{j,\,k},&
		\end{alignedat}
	\end{equation}
	and straightforward calculation yields $I_\Phi(\ket{\text{SRU}_2(\pi, \Phi;\mu)}) = 8 S_M$. Thus, we propose another Ansatz of quarter-turn, $\gamma = \pi/2$, which yields the expression for the derivative,
	
	\begin{equation}
		\begin{alignedat}{3}
			&\ket{\partial_\Phi\text{SRU}_2\qty(\frac{\pi}{2},\Phi;\mu)}= && \frac{i }{2^{S_M + S_P}}\sum_{j=-S_M}^{S_M} \sum_{k = -S_P}^{S_P}
			\sqrt{\binom{2S_M}{S_M - j}\binom{2S_P}{S_P - k}}
			& \\
			&&& \times\qty(
			\frac{S_M + j}{e^{-i(\Phi -  k \mu )} + 1} + 
			\frac{S_M - j}{e^{i (\Phi-k\mu) }-1}) \qty(1 + e^{i(\Phi-k \mu)})^{S_M + j}  & \\
			&&& \times\qty(1 - e^{-i(\Phi-k \mu)})^{S_M - j}  \ket{j,\,k}.&
		\end{alignedat}
	\end{equation}
	Making use of the sums defined in \eqref{eq:F_expressions} and similar scalar products as in \eqref{eq:F_products} we arrive at the expression \eqref{eq:axis_estimation_QFI} for the QFI for axis angle estimation.

	\section{Quantum Fisher information for superchannel estimation} \label{app:superchannel_estim}

	Recall the initial setting of the superchannel estimation problem,
	
	\begin{align}
		\ket{\psi_\gamma} = e^{i\gamma H_0} U_1 e^{-i\gamma H_0}\ket{\psi_0}, &&
		\partial_\gamma \ket{\psi_\gamma} = i \comm{H_0}{e^{i\gamma H_0} U_1 e^{-i\gamma H_0}}\ket{\psi_0},
	\end{align}
	where we assume $H_0$ to be a fixed Hamiltonian with the maximal and minimal eigenvalues given by $\lambda_+ = -\lambda_- = \frac{N}{2}$ and $\ket{\psi_0}$ and $U_1 = e^{i H_1}$ as free quantum state and a unitary with a free Hamiltonian $H_1$, respectively, to be adjusted for maximal QFI for the estimation of $\gamma$. In order to demonstrate the lower bound on this maximum we consider $\ket{\psi_0} = \frac{1}{\sqrt{2}}\qty(\ket{\lambda_+} + e^{i\alpha}\ket{\lambda_-})$ as an equal superposition of the eigenstates related to the extremal eigenvalues. With this assumption we evaluate the inner products relevant to the QFI,
		\begin{align}
			\braket{\psi_\gamma}{\partial_\gamma \psi_\gamma} = & i\qty(\ev{e^{i\gamma H_0}U_1^{\dagger}H_0U_1e^{-i\gamma H_0}}{\psi_0} - \ev{H_0}{\psi_0}) \\
			= & \frac{i}{2}\left(\ev{U_1^{\dagger}H_0U_1}{\lambda_-} + \ev{U_1^{\dagger}H_0U_1}{\lambda_+}\right. \nonumber\\
			& + e^{-i(\alpha + N\gamma)}\mel{\lambda_-}{U_1^{\dagger}H_0U_1}{\lambda_+} + e^{i(\alpha + N\gamma)}\mel{\lambda_+}{U_1^{\dagger}H_0U_1}{\lambda_-}\left.\right) \\
			= & \frac{i}{2}\ev{U_1^{\dagger}H_0U_1}{\lambda_+ + e^{i(\alpha + N\gamma)} \lambda_-},
		\end{align}
		and a similar array of expression for the other product of interest,
		\begin{align}
			\braket{\partial_\gamma\psi_\gamma} = \bra{\psi_0}& \qty(H_0 e^{i\gamma H_0} U_1 e^{-i\gamma H_0} - e^{i\gamma H_0} U_1 e^{-i\gamma H_0} H_0 ) \nonumber\\
			& \times\qty(e^{i\gamma H_0} U_1^{\dagger} e^{-i\gamma H_0}H_0  -  H_0 e^{i\gamma H_0} U_1^{\dagger} e^{-i\gamma H_0}) \ket{\psi_0} \\
			= \bra{\psi_0}&\left(H_0^2 + e^{i\gamma H_0} U_1^{\dagger} H_0^2 U_1 e^{-i\gamma H_0}\right. \nonumber \\
			 - & \left. e^{i\gamma H_0} U_1^{\dagger} H_0 U_1 e^{-i\gamma H_0} H_0 -  H_0 e^{i\gamma H_0} U_1^{\dagger} H_0 U_1 e^{-i\gamma H_0}\right)\ket{\psi_0}.
		\end{align}
		At this point we decompose $\ket{\psi_0}$ in terms of eigenvectors of $H_0$,
		\begin{align}
			\braket{\partial_\gamma\psi_\gamma} = &\frac{N^2}{4} + \frac{1}{2}\ev{U_1^{\dagger} H_0^2 U_1}{\lambda_+ + e^{i(\alpha+N\gamma)}\lambda_-}\\
			& - \frac{N}{2}\ev{U_1^{\dagger} H_0 U_1}{\lambda_+}
			+  \frac{N}{2} \ev{U_1^{\dagger} H_0 U_1}{\lambda_-}.
		\end{align}
		Observe that $H_0 = \frac{N}{2}\sigma_z$ acts on the subsapce spanned by $\qty{\ket{\lambda_-},\ket{\lambda_+}}$, so it can yield only smaller contributions by mixing with the remaining eigenspaces. Assuming that $U_1$ acts only within the same subspace as $H_0$ allows us to write
		\begin{align}
			\ket{0} \equiv \ket{\lambda_+}, &&
			\ket{1} \equiv \ket{\lambda_-}, &&
			U_1 = \cos\theta\,\mathbb{I} + i \sin\theta\,\sigma_{\va{k}}.
		\end{align}
		In this way we arrive at the expressions,
		\begin{align}
			\braket{\partial_\gamma\psi_\gamma} & = \frac{N^2}{2} - \frac{N^2}{4}\Tr[\sigma_z\qty(\cos\theta\,\mathbb{I} - i \sin\theta\, \sigma_{\va{k}})\sigma_z\qty(\cos\theta\,\mathbb{I} + i \sin\theta\, \sigma_{\va{k}})] \\
			\braket{\psi_\gamma}{\partial_\gamma\psi_\gamma} 
			& = i\frac{N}{4}\sin(2\theta) \ev{\qty(\sigma_z\sigma_{\va{k}} - \sigma_{\va{k}}\sigma_z)}{0 + e^{i(\alpha+N\gamma)}1}
		\end{align}
		which, by setting $\va{k} = \cos \phi \vu{x} + \sin \phi \vu{y}$ and $\theta = \pi/2$ yield,
		\begin{equation}
			\braket{\partial_\gamma\psi_\gamma} = \frac{N^2}{2} - \frac{N^2}{4}\Tr[\sigma_z \sigma_{\va{k}}\sigma_z\sigma_{\va{k}}] = \frac{N^2}{2} + \frac{N^2}{4}\Tr[\sigma^2_{\va{k}_\perp}] = N^2,
		\end{equation}
		and $\braket{\psi_\gamma}{\partial_\gamma\psi_\gamma} = 0$. Therefore, we find the QFI to be equal to
		\begin{equation}
			\max_{\ket{\psi_0},U_1} I(\ket{\psi_\gamma};\gamma) =  \max_{\ket{\psi_0},U_1}4\qty(\braket{\partial_\gamma\psi_\gamma} - \abs{\braket{\psi_\gamma}{\partial_\gamma\psi_\gamma}}^2)  \geq 4N^2,
		\end{equation}
		which is higher than the value of $\max_\Phi I_\Phi(\ket{\text{SRU}_2(\sfrac{\pi}{2},\Phi;\pi)}) = N^2 + N$ found for the axis estimation problem. 
		In order to see that $4N^2$ is indeed the maximum achievable value consider a projection operator $P = \op{\lambda_+} + \op{\lambda_-}$ onto the space connected to the largest and smallest eigenvalues. Within this subspace the $U_1$ indeed acts as a time-reversal operation on one of the imprinting protocols,
		\begin{equation}
			Pe^{i\gamma H_0} U_1 e^{-i\gamma H_0}\ket{\psi_0}P = Pe^{2i\gamma H_0} P,
		\end{equation}
		and by this virtue the value is reconnected to the standard Hamiltonian estimation problem with doubled time of action.
	\pagebreak
	\vspace{10cm}
	\normalem
	\bibliographystyle{quantum_abbr}
	\bibliography{biblio}

\begin{thebibliography}{10}

\bibitem{DBS13}
R.~Demkowicz-Dobrza{\'n}ski, K.~Banaszek, and R.~Schnabel.
\newblock ``Fundamental quantum interferometry bound for the
  squeezed-light-enhanced gravitational wave detector {GEO} 600''.
\newblock
  \href{https://dx.doi.org/https://dx.doi.org/10.1103/PhysRevA.88.041802}{Phys.
  Rev. A {\bf 88}, 041802}~(2013).

\bibitem{AEtAl13}
J.~Aasi et~al.
\newblock ``Enhanced sensitivity of the {LIGO} gravitational wave detector by
  using squeezed states of light''.
\newblock \href{https://dx.doi.org/10.1038/nphoton.2013.177}{Nature
  Photonics}~(2013).

\bibitem{IR18}
I.~G. Irastorza and J.~Redondo.
\newblock ``New experimental approaches in the search for axion-like
  particles''.
\newblock
  \href{https://dx.doi.org/https://doi.org/10.1016/j.ppnp.2018.05.003}{Prog.
  Part. Nucl. Phys. {\bf 102}, 89--159}~(2018).

\bibitem{BEtAl19}
S.~C. Burd, R.~Srinivas, J.~J. Bollinger, A.~C. Wilson, D.~J. Wineland,
  D.~Leibfried, D.~H. Slichter, and D.~T.~C. Allcock.
\newblock ``Quantum amplification of mechanical oscillator motion''.
\newblock \href{https://dx.doi.org/10.1126/science.aaw2884}{Science {\bf 364},
  1163--1165}~(2019).

\bibitem{YMK86}
B.~Yurke, S.~L. McCall, and J.~R. Klauder.
\newblock ``{SU}(2) and {SU}(1,1) interferometers''.
\newblock \href{https://dx.doi.org/10.1103/PhysRevA.33.4033}{Phys. Rev. A {\bf
  33}, 4033--4054}~(1986).

\bibitem{Caves20}
C.~M. Caves.
\newblock ``Reframing {SU}(1,1) interferometry''.
\newblock \href{https://dx.doi.org/https://doi.org/10.1002/qute.201900138}{Adv.
  Quantum Technol. {\bf 3}, 1900138}~(2020).

\bibitem{PhysRevX.2.031016}
M.~Tsang and C.~M. Caves.
\newblock ``Evading quantum mechanics: Engineering a classical subsystem within
  a quantum environment''.
\newblock \href{https://dx.doi.org/10.1103/PhysRevX.2.031016}{Phys. Rev. X {\bf
  2}, 031016}~(2012).

\bibitem{AgarwalQOpticsBook}
G.~Agarwal.
\newblock ``Quantum optics''.
\newblock Quantum Optics. Cambridge University Press. ~(2013).

\bibitem{bergeal_phase-preserving_2010}
N.~Bergeal, F.~Schackert, M.~Metcalfe, R.~Vijay, V.~E. Manucharyan, L.~Frunzio,
  D.~E. Prober, R.~J. Schoelkopf, S.~M. Girvin, and M.~H. Devoret.
\newblock ``Phase-preserving amplification near the quantum limit with a
  {Josephson} ring modulator''.
\newblock \href{https://dx.doi.org/10.1038/nature09035}{Nature {\bf 465},
  64--68}~(2010).

\bibitem{PhysRevB.98.045405}
A.~Roy and M.~Devoret.
\newblock ``Quantum-limited parametric amplification with {Josephson} circuits
  in the regime of pump depletion''.
\newblock \href{https://dx.doi.org/10.1103/PhysRevB.98.045405}{Phys. Rev. B
  {\bf 98}, 045405}~(2018).

\bibitem{PhysRevLett.106.110502}
R.~Vijay, D.~H. Slichter, and I.~Siddiqi.
\newblock ``Observation of quantum jumps in a superconducting artificial
  atom''.
\newblock \href{https://dx.doi.org/10.1103/PhysRevLett.106.110502}{Phys. Rev.
  Lett. {\bf 106}, 110502}~(2011).

\bibitem{renger_beyond_2021}
M.~Renger, S.~Pogorzalek, Q.~Chen, Y.~Nojiri, K.~Inomata, Y.~Nakamura,
  M.~Partanen, A.~Marx, R.~Gross, F.~Deppe, and K.~G. Fedorov.
\newblock ``Beyond the standard quantum limit for parametric amplification of
  broadband signals''.
\newblock \href{https://dx.doi.org/10.1038/s41534-021-00495-y}{npj Quantum Inf.
  {\bf 7}, 1--7}~(2021).

\bibitem{PhysRevLett.118.103601}
C.~F. Ockeloen-Korppi, E.~Damsk\"agg, J.-M. Pirkkalainen, T.~T. Heikkil\"a,
  F.~Massel, and M.~A. Sillanp\"a\"a.
\newblock ``Noiseless quantum measurement and squeezing of microwave fields
  utilizing mechanical vibrations''.
\newblock \href{https://dx.doi.org/10.1103/PhysRevLett.118.103601}{Phys. Rev.
  Lett. {\bf 118}, 103601}~(2017).

\bibitem{Sik83}
P.~Sikivie.
\newblock ``Experimental tests of the "invisible" axion''.
\newblock \href{https://dx.doi.org/10.1103/PhysRevLett.51.1415}{Phys. Rev.
  Lett. {\bf 51}, 1415--1417}~(1983).

\bibitem{zheng_accelerating_2016}
H.~Zheng, M.~Silveri, R.~T. Brierley, S.~M. Girvin, and K.~W. Lehnert.
\newblock ``Accelerating dark-matter axion searches with quantum measurement
  technology''~(2016).
\newblock arXiv:1607.02529.

\bibitem{KU93}
M.~Kitagawa and M.~Ueda.
\newblock ``Squeezed spin states''.
\newblock \href{https://dx.doi.org/10.1103/PhysRevA.47.5138}{Phys. Rev. A {\bf
  47}, 5138--5143}~(1993).

\bibitem{Byr13}
T.~Byrnes.
\newblock ``Fractality and macroscopic entanglement in two-component
  {B}ose-{E}instein condensates''.
\newblock \href{https://dx.doi.org/10.1103/PhysRevA.88.023609}{Phys. Rev. A
  {\bf 88}, 023609}~(2013).

\bibitem{KPST13}
H.~Kurkjian, K.~Paw\l{}owski, A.~Sinatra, and P.~Treutlein.
\newblock ``Spin squeezing and {E}instein-{P}odolsky-{R}osen entanglement of
  two bimodal condensates in state-dependent potentials''.
\newblock \href{https://dx.doi.org/10.1103/PhysRevA.88.043605}{Phys. Rev. A
  {\bf 88}, 043605}~(2013).

\bibitem{SDC11}
A.~Sinatra, J.-C. Dornstetter, and Y.~Castin.
\newblock ``Spin squeezing in {B}ose-{E}instein condensates: Limits imposed by
  decoherence and non-zero temperature''.
\newblock \href{https://dx.doi.org/10.1007/s11467-011-0219-7}{Front. Phys. {\bf
  7}, 86--97}~(2011).

\bibitem{CGSP21}
R.~Corgier, N.~Gaaloul, A.~Smerzi, and L.~Pezz\`e.
\newblock ``Delta-kick squeezing''.
\newblock \href{https://dx.doi.org/10.1103/PhysRevLett.127.183401}{Phys. Rev.
  Lett. {\bf 127}, 183401}~(2021).

\bibitem{BEtAl21}
T.~Bilitewski, L.~De~Marco, J.-R. Li, K.~Matsuda, W.~G. Tobias, G.~Valtolina,
  J.~Ye, and A.~M. Rey.
\newblock ``Dynamical generation of spin squeezing in ultracold dipolar
  molecules''.
\newblock \href{https://dx.doi.org/10.1103/PhysRevLett.126.113401}{Phys. Rev.
  Lett. {\bf 126}, 113401}~(2021).

\bibitem{LEtAl04}
D.~Leibfried, M.~D. Barrett, T.~Schaetz, J.~Britton, J.~Chiaverini, W.~M.
  Itano, J.~D. Jost, C.~Langer, and D.~J. Wineland.
\newblock ``Toward {H}eisenberg-limited spectroscopy with multiparticle
  entangled states''.
\newblock \href{https://dx.doi.org/10.1126/science.1097576}{Science {\bf 304},
  1476--1478}~(2004).

\bibitem{CEtAl21}
C.~Chryssomalakos, L.~Hanotel, E.~Guzm\'an-Gonz\'alez, D.~Braun,
  E.~Serrano-Ens\'astiga, and K.~\ifmmode~\dot{Z}\else \.{Z}\fi{}yczkowski.
\newblock ``Symmetric multiqudit states: Stars, entanglement, and
  rotosensors''.
\newblock \href{https://dx.doi.org/10.1103/PhysRevA.104.012407}{Phys. Rev. A
  {\bf 104}, 012407}~(2021).

\bibitem{KEtAl19}
R.~Kaubruegger, P.~Silvi, C.~Kokail, R.~van Bijnen, A.~M. Rey, J.~Ye, A.~M.
  Kaufman, and P.~Zoller.
\newblock ``Variational spin-squeezing algorithms on programmable quantum
  sensors''.
\newblock \href{https://dx.doi.org/10.1103/PhysRevLett.123.260505}{Phys. Rev.
  Lett. {\bf 123}, 260505}~(2019).

\bibitem{REtAl10}
M.~F. Riedel, P.~Böhi, Y.~Li, T.~W. Hänsch, A.~Sinatra, and P.~Treutlein.
\newblock ``Atom-chip-based generation of entanglement for quantum metrology''.
\newblock \href{https://dx.doi.org/10.1038/nature08988}{Nature}~(2010).

\bibitem{CEtAl22}
S.~Colombo, E.~Pedrozo-Peñafiel, A.~F. {Adiyatullin na nAff}, Z.~Li,
  E.~Mendez, C.~Shu, and V.~Vuletić.
\newblock ``Time-reversal-based quantum metrology with many-body entangled
  states''.
\newblock \href{https://dx.doi.org/10.1038/s41567-022-01653-5}{Nature
  Physics}~(2022).

\bibitem{KSVZ23}
R.~Kaubruegger, A.~Shankar, D.~V. Vasilyev, and P.~Zoller.
\newblock ``Optimal and variational multi-parameter quantum metrology and
  vector field sensing''.
\newblock arXiv preprint: 2302.07785~(2023).

\bibitem{Rad71}
J.~M. Radcliffe.
\newblock ``Some properties of coherent spin states''.
\newblock \href{https://dx.doi.org/10.1088/0305-4470/4/3/009}{J. Phys. A: Gen.
  Phys. {\bf 4}, 313--323}~(1971).

\bibitem{ACGT72}
F.~T. Arecchi, E.~Courtens, R.~Gilmore, and H.~Thomas.
\newblock ``Atomic coherent states in quantum optics''.
\newblock \href{https://dx.doi.org/10.1103/PhysRevA.6.2211}{Phys. Rev. A {\bf
  6}, 2211--2237}~(1972).

\bibitem{HH74}
R.~Holtz and J.~Hanus.
\newblock ``On coherent spin states''.
\newblock \href{https://dx.doi.org/10.1088/0305-4470/7/4/001}{J. Phys. A: Math.
  Nucl. Gen. {\bf 7}, L37--L40}~(1974).

\bibitem{ZFG90}
W.-M. Zhang, D.~H. Feng, and R.~Gilmore.
\newblock ``Coherent states: Theory and some applications''.
\newblock \href{https://dx.doi.org/10.1103/RevModPhys.62.867}{Rev. Mod. Phys.
  {\bf 62}, 867--927}~(1990).

\bibitem{Agar81}
G.~S. Agarwal.
\newblock ``Relation between atomic coherent-state representation, state
  multipoles, and generalized phase-space distributions''.
\newblock \href{https://dx.doi.org/10.1103/PhysRevA.24.2889}{Phys. Rev. A {\bf
  24}, 2889--2896}~(1981).

\bibitem{DAS94}
J.~P. Dowling, G.~S. Agarwal, and W.~P. Schleich.
\newblock ``Wigner distribution of a general angular-momentum state:
  Applications to a collection of two-level atoms''.
\newblock \href{https://dx.doi.org/10.1103/PhysRevA.49.4101}{Phys. Rev. A {\bf
  49}, 4101--4109}~(1994).

\bibitem{DKMG21}
J.~Davis, M.~Kumari, R.~B. Mann, and S.~Ghose.
\newblock ``Wigner negativity in spin-$j$ systems''.
\newblock \href{https://dx.doi.org/10.1103/PhysRevResearch.3.033134}{Phys. Rev.
  Research {\bf 3}, 033134}~(2021).

\bibitem{KZ04}
A.~Kenfack and K.~Życzkowski.
\newblock ``Negativity of the {W}igner function as an indicator of
  non-classicality''.
\newblock \href{https://dx.doi.org/10.1088/1464-4266/6/10/003}{J. Opt., B
  Quantum semiclass. {\bf 6}, 396}~(2004).

\bibitem{KEtAl20}
J.~Kitzinger, M.~Chaudhary, M.~Kondappan, V.~Ivannikov, and T.~Byrnes.
\newblock ``Two-axis two-spin squeezed states''.
\newblock \href{https://dx.doi.org/10.1103/PhysRevResearch.2.033504}{Phys. Rev.
  Research {\bf 2}, 033504}~(2020).

\bibitem{Hels67}
C.~Helstrom.
\newblock ``Minimum mean-squared error of estimates in quantum statistics''.
\newblock
  \href{https://dx.doi.org/https://doi.org/10.1016/0375-9601(67)90366-0}{Phys.
  Lett. A {\bf 25}, 101--102}~(1967).

\bibitem{braunstein_statistical_1994}
S.~L. Braunstein and C.~M. Caves.
\newblock ``Statistical distance and the geometry of quantum states''.
\newblock \href{https://dx.doi.org/10.1103/PhysRevLett.72.3439}{Phys. Rev.
  Lett. {\bf 72}, 3439}~(1994).

\bibitem{braunstein_generalized_1996}
S.~L. Braunstein, C.~M. Caves, and G.~J. Milburn.
\newblock ``Generalized uncertainty relations: {Theory}, examples, and
  {Lorentz} invariance''.
\newblock \href{https://dx.doi.org/10.1006/aphy.1996.0040}{Annals of Physics
  {\bf 247}, 135--173}~(1996).

\bibitem{Paris09}
M.~G.~A. Paris.
\newblock ``Quantum estimation for quantum technology''.
\newblock \href{https://dx.doi.org/}{Int. J. Quantum Inf. {\bf 7}, 125}~(2009).

\bibitem{Fraise17}
J.~M.~E. Fraïsse.
\newblock ``New concepts in quantum-metrology: {F}rom coherent averaging to
  {H}amiltonian extensions''.
\newblock PhD thesis.
\newblock University of T{\"u}bingen.
\newblock ~(2017).

\bibitem{LYLW19}
J.~Liu, H.~Yuan, X.-M. Lu, and X.~Wang.
\newblock ``Quantum {Fisher} information matrix and multiparameter
  estimation''.
\newblock \href{https://dx.doi.org/10.1088/1751-8121/ab5d4d}{J. Phys. A Math.
  Theor.}~(2019).

\bibitem{Giraud10}
O.~Giraud, P.~Braun, and D.~Braun.
\newblock ``Quantifying quantumness and the quest for queens of quantum''.
\newblock \href{https://dx.doi.org/10.1088/1367-2630/12/6/063005}{New J. Phys.
  {\bf 12}, 063005}~(2010).

\bibitem{H69}
C.~W. Helstrom.
\newblock ``Quantum detection and estimation theory''.
\newblock \href{https://dx.doi.org/10.1007/bf01007479}{J. Stat. Phys. {\bf 1},
  231--252}~(1969).

\bibitem{H11}
A.~Holevo.
\newblock ``Probabilistic and statistical aspects of quantum theory''.
\newblock \href{https://dx.doi.org/10.1007/978-88-7642-378-9}{Edizioni della
  Normale Pisa}. ~(2011).

\bibitem{demkowicz-dobrzanski_multi-parameter_2020}
R.~Demkowicz-Dobrza{\'n}ski, W.~G{\'o}recki, and M.~Gu{\c{t}}{\u{a}}.
\newblock ``Multi-parameter estimation beyond {Quantum} {Fisher}
  {Information}''.
\newblock \href{https://dx.doi.org/10.1088/1751-8121/ab8ef3}{J. Phys. A: Math.
  Theor. {\bf 53}, 363001}~(2020).

\bibitem{holevo_statistical_1973}
A.~S. Holevo.
\newblock ``Statistical decision theory for quantum systems''.
\newblock \href{https://dx.doi.org/10.1016/0047-259X(73)90028-6}{J. Multivar.
  Anal. {\bf 3}, 337--394}~(1973).

\bibitem{RJD16}
S.~Ragy, M.~Jarzyna, and R.~Demkowicz-Dobrza{\'n}ski.
\newblock ``Compatibility in multiparameter quantum metrology''.
\newblock \href{https://dx.doi.org/10.1103/PhysRevA.94.052108}{Phys. Rev. A
  {\bf 94}, 052108}~(2016).

\bibitem{AFD19}
F.~Albarelli, J.~F. Friel, and A.~Datta.
\newblock ``Evaluating the {Holevo} {Cram\'er}-{Rao} bound for multiparameter
  quantum metrology''.
\newblock \href{https://dx.doi.org/10.1103/PhysRevLett.123.200503}{Phys. Rev.
  Lett. {\bf 123}, 200503}~(2019).

\bibitem{tsang_quantum_2020}
M.~Tsang, F.~Albarelli, and A.~Datta.
\newblock ``Quantum {{Semiparametric Estimation}}''.
\newblock \href{https://dx.doi.org/10.1103/PhysRevX.10.031023}{Phys. Rev. X
  {\bf 10}, 031023}~(2020).

\bibitem{WBY85}
F.~Waldner, D.~R. Barberis, and H.~Yamazaki.
\newblock ``Route to chaos by irregular periods: Simulations of parallel
  pumping in ferromagnets''.
\newblock \href{https://dx.doi.org/10.1103/PhysRevA.31.420}{Phys. Rev. A {\bf
  31}, 420--431}~(1985).

\bibitem{KSH87}
M.~Kuś, R.~Scharf, and F.~Haake.
\newblock ``Symmetry versus degree of level repulsion for kicked quantum
  systems''.
\newblock \href{https://dx.doi.org/10.1007/bf01312770}{Z. Phys. B: Condens.
  Matter {\bf 66}, 129--134}~(1987).

\bibitem{HS88}
F.~Haake and D.~L. Shepelyansky.
\newblock ``The kicked rotator as a limit of the kicked top''.
\newblock \href{https://dx.doi.org/10.1209/0295-5075/5/8/001}{Europhys. Lett.
  {\bf 5}, 671--676}~(1988).

\bibitem{BGHS96}
P.~A. Braun, P.~Gerwinski, F.~Haake, and H.~Schomerus.
\newblock ``Semiclassics of rotation and torsion''.
\newblock \href{https://dx.doi.org/10.1007/s002570050101}{Z. Phys. B: Condens.
  Matter {\bf 100}, 115--127}~(1996).

\bibitem{Metal23}
G.~Müller-Rigat, A.~K. Srivastava, S.~Kurdziałek, G.~Rajchel-Mieldzioć,
  M.~Lewenstein, and I.~Frérot.
\newblock ``Certifying the quantum fisher information from a given set of mean
  values: a semidefinite programming approach''~(2023).
\newblock
  url:~\href{https://arxiv.org/abs/2306.12711v2}{arxiv.org/abs/2306.12711v2}.

\bibitem{Z01}
W.~H. Zurek.
\newblock ``Sub-{Planck} structure in phase space and its relevance for quantum
  decoherence''.
\newblock \href{https://dx.doi.org/10.1038/35089017}{Nature}~(2001).

\bibitem{JD12}
M.~Jarzyna and R.~Demkowicz-Dobrza\ifmmode~\acute{n}\else \'{n}\fi{}ski.
\newblock ``Quantum interferometry with and without an external phase
  reference''.
\newblock \href{https://dx.doi.org/10.1103/PhysRevA.85.011801}{Phys. Rev. A
  {\bf 85}, 011801}~(2012).

\bibitem{YEtAl19}
C.~You, S.~Adhikari, X.~Ma, M.~Sasaki, M.~Takeoka, and J.~P. Dowling.
\newblock ``Conclusive precision bounds for {SU}(1,1) interferometers''.
\newblock \href{https://dx.doi.org/10.1103/PhysRevA.99.042122}{Phys. Rev. A
  {\bf 99}, 042122}~(2019).

\bibitem{aharonov_quantum_1984}
Y.~Aharonov and T.~Kaufherr.
\newblock ``Quantum frames of reference''.
\newblock \href{https://dx.doi.org/10.1103/PhysRevD.30.368}{Phys. Rev. D {\bf
  30}, 368--385}~(1984).

\bibitem{bartlett_reference_2007}
S.~D. Bartlett, T.~Rudolph, and R.~W. Spekkens.
\newblock ``Reference frames, superselection rules, and quantum information''.
\newblock \href{https://dx.doi.org/10.1103/RevModPhys.79.555}{Rev. Mod. Phys.
  {\bf 79}, 555}~(2007).

\bibitem{giacomini_quantum_2022}
F.~Giacomini and C.~Brukner.
\newblock ``Quantum superposition of spacetimes obeys {Einstein}'s equivalence
  principle''.
\newblock \href{https://dx.doi.org/10.1116/5.0070018}{AVS Quantum Science {\bf
  4}, 015601}~(2022).

\bibitem{KW15}
D.~Kajtoch and E.~Witkowska.
\newblock ``Quantum dynamics generated by the two-axis countertwisting
  {Hamiltonian}''.
\newblock \href{https://dx.doi.org/10.1103/PhysRevA.92.013623}{Phys. Rev. A
  {\bf 92}, 013623}~(2015).

\bibitem{HEtAl22}
T.~Hern\'andez~Yanes, M.~P\l{}odzie\ifmmode~\acute{n}\else \'{n}\fi{},
  M.~Mackoit Sinkevi\ifmmode \check{c}\else \v{c}\fi{}ien\ifmmode~\dot{e}\else
  \.{e}\fi{}, G.~\ifmmode~\check{Z}\else \v{Z}\fi{}labys,
  G.~Juzeli\ifmmode~\bar{u}\else \={u}\fi{}nas, and E.~Witkowska.
\newblock ``One- and two-axis squeezing via laser coupling in an atomic
  {Fermi}-{Hubbard} model''.
\newblock \href{https://dx.doi.org/10.1103/PhysRevLett.129.090403}{Phys. Rev.
  Lett. {\bf 129}, 090403}~(2022).

\bibitem{LW21}
X.-M. Lu and X.~Wang.
\newblock ``Incorporating {Heisenberg's} uncertainty principle into quantum
  multiparameter estimation''.
\newblock \href{https://dx.doi.org/10.1103/PhysRevLett.126.120503}{Phys. Rev.
  Lett. {\bf 126}, 120503}~(2021).

\bibitem{ferretti2023generating}
H.~Ferretti, Y.~B. Yilmaz, K.~Bonsma-Fisher, A.~Z. Goldberg, N.~Lupu-Gladstein,
  A.~O.~T. Pang, L.~A. Rozema, and A.~M. Steinberg.
\newblock ``Generating a 4-photon tetrahedron state: Towards simultaneous
  super-sensitivity to non-commuting rotations''~(2023).
\newblock
  url:~\href{https://arxiv.org/abs/2310.17150v1}{arxiv.org/abs/2310.17150v1}.

\bibitem{chirkin_1993}
A.~S. Chirkin, A.~A. Orlov, and D.~Y. Parashchuk.
\newblock ``Quantum theory of two-mode interactions in optically anisotropic
  media with cubic nonlinearities: Generation of quadrature- and
  polarization-squeezed light''.
\newblock \href{https://dx.doi.org/10.1070/QE1993v023n10ABEH003182}{Quantum
  Elec. {\bf 23}, 870}~(1993).

\bibitem{klyshko97polarization}
D.~M. Klyshko.
\newblock ``Polarization of light: Fourth-order effects and
  polarization-squeezed states''.
\newblock \href{https://dx.doi.org/10.1134/1.558243}{JETP {\bf 84},
  1065--1079}~(1997).

\bibitem{korolkova02}
N.~Korolkova, G.~Leuchs, R.~Loudon, T.~C. Ralph, and C.~Silberhorn.
\newblock ``Polarization squeezing and continuous-variable polarization
  entanglement''.
\newblock \href{https://dx.doi.org/10.1103/PhysRevA.65.052306}{Phys. Rev. A
  {\bf 65}, 052306}~(2002).

\bibitem{schnabel03}
R.~Schnabel, W.~P. Bowen, N.~Treps, T.~C. Ralph, H.-A. Bachor, and P.~K. Lam.
\newblock ``Stokes-operator-squeezed continuous-variable polarization states''.
\newblock \href{https://dx.doi.org/10.1103/PhysRevA.67.012316}{Phys. Rev. A
  {\bf 67}, 012316}~(2003).

\end{thebibliography}
	
\end{document}